\documentclass[aip,prl,reprint,showpacs,amsmath,amssymb]{revtex4-1}
\pdfoutput=1 

\usepackage{tikz}
\usepackage{graphicx}
\usepackage{dcolumn}
\usepackage{bm}
\usepackage{upgreek}
\usepackage{array}
\usepackage{booktabs}
\usepackage{microtype}
\usepackage{cmap}
\usepackage{geometry}
\usepackage{siunitx}

\usepackage{tabularx}
\usepackage{multirow}
\usepackage{chngpage}

\DeclareSIUnit{\cal}{cal}
\DeclareSIUnit\Molar{\textsc{m}}

\usepackage{hyperref}
\hypersetup{colorlinks, citecolor={blue}, linkcolor={red}, urlcolor={violet}, pdftitle={DNA hairpins destabilize duplexes primarily by promoting melting rather than by inhibiting hybridization}, pdfauthor={J. S. Schreck, T. E. Ouldridge, F. Romano, P. {\v{S}}ulc, L. Shaw, A. A. Louis, J. P. K. Doye}, pdfdisplaydoctitle}

\begin{document}

\author{John S. Schreck}
\affiliation{Physical and Theoretical Chemistry Laboratory, Department of Chemistry, University of Oxford, South Parks Road, Oxford OX1 3QZ, United Kingdom}

\author{Thomas E. Ouldridge}
\affiliation{Rudolph Peierls Centre for Theoretical Physics, University of Oxford, 1 Keble Road, Oxford OX1 3NP, United Kingdom}

\author{Flavio Romano}
\affiliation{Physical and Theoretical Chemistry Laboratory, Department of Chemistry, University of Oxford, South Parks Road, Oxford OX1 3QZ, United Kingdom}

\author{Petr {\v{S}}ulc}
\affiliation{Rudolph Peierls Centre for Theoretical Physics, University of Oxford, 1 Keble Road, Oxford OX1 3NP, United Kingdom}

\author{Liam Shaw}
\affiliation{Rudolph Peierls Centre for Theoretical Physics, University of Oxford, 1 Keble Road, Oxford OX1 3NP, United Kingdom}

\author{Ard A. Louis}
\affiliation{Rudolph Peierls Centre for Theoretical Physics,  University of Oxford, 1 Keble Road, Oxford OX1 3NP, United Kingdom}

\author{Jonathan P. K. Doye}
\affiliation{Physical and Theoretical Chemistry Laboratory, Department of Chemistry, University of Oxford, South Parks Road, Oxford OX1 3QZ, United Kingdom}

\title{DNA hairpins primarily promote duplex melting rather than inhibiting hybridization}

\date{\today}

\begin{abstract}
The effect of secondary structure on DNA duplex formation is poorly understood. We use a coarse-grained model of DNA to show that specific 3- and 4-base pair hairpins reduce hybridization rates by factors of 2 and 10 respectively, in good agreement with experiment. By contrast, melting rates are accelerated by factors of $\sim$100 and $\sim$2000. This surprisingly large speed-up occurs because hairpins form during the melting process, stabilizing partially melted states, and facilitating dissociation. These results may help guide the design of DNA devices that use hairpins to modulate hybridization and dissociation pathways and rates.
\end{abstract}

\maketitle

The field of DNA nanotechnology has grown enormously since Seeman's original work in the early 1980s, which suggested that the specificity of DNA hybridization could be harnessed to permit the design of artificial structures \cite{Seeman1982}. Large scale structures can now be designed to self-assemble with nanoscale precision \cite{Rothemund06, Zheng2009, Dietz2009, Ke2012}. DNA based switches and motors have been demonstrated~\cite{Yurke2000,Venkataraman2007,Bath2005,Bath2009,Omabegho2009}, and decision-making constructs have been shown to interact with biological systems~\cite{Douglas2012}. Other work has explored the potential for DNA-based computation~\cite{Seelig2006, Qian2011, Chen2013}. The fundamental ingredient in the self-assembly of DNA nanostructures and in the operation of DNA nanomachines is the hybridization of single-stranded DNA to form duplexes.

Besides the canonical double-helical duplex, DNA hairpins, in which a self-complementary strand loops around to bind to itself, are perhaps the simplest structure that DNA can form. In nanotechnological applications, hairpins have the potential to be both advantageous and deleterious. For example, metastable hairpins may impede hybridization by blocking binding sites, thereby providing an additional barrier to hybridization~\cite{Gao2006}. This can be a nuisance for DNA nanostructures that are assembled at low temperature, especially those structures that are composed of longer strands, since such single strands are likely to possess intra-strand bonds that will inhibit assembly~\cite{Romano_overstretch_2013}. On the other hand, hairpins can also be used constructively to control some features of reaction pathways. For example, metastable hairpins can be a source of fuel for autonomous DNA machines~\cite{Turberfield2003,Venkataraman2007,Green2008,Omabegho2009}, and can be used to suppress undesirable leak reactions~\cite{Tomov2013}. 

Despite the importance of hairpins to DNA nanotechnology, a systematic understanding of their influence on the hybridization transition  remains elusive. Recently, however, an important step in this direction was made by Gao {\it et al.} who measured the rate of hybridization at 20$^\circ$C and high salt for several systems of complementary strands, where the strands contained either no stable hairpins, or stable 3- or 4-base pair hairpins~\cite{Gao2006}. The hybridization rates of the 3-base pair hairpins were reduced by a factor of approximately 2, and those for the 4-base pair hairpins by an order of magnitude. While these are not negligible effects, they are smaller than might na\"{i}vely be expected. Assuming a second-order duplex formation transition, the hybridization ($k_+$) and melting ($k_-$) rate constants are necessarily related by the free-energy change of duplex formation $\Delta G^0$ with respect to the single-stranded state through
\begin{equation}
\frac{k_+}{k_-} = \frac{\exp (-\Delta G^0/RT)}{c^0},
\label{equ1}
\end{equation}
where $c^0$ = \SI{1}{\Molar}. Gao {\it et al.} found that both the 3- and 4-base pair hairpins were very
stable at 20$^\circ$C~\cite{Gao2006}, suggesting a significant change in
$\Delta G^0$ relative to the hairpin-free system (particularly in the 4-base
pair case). However, the measured reduction in the hybridization rate $k_+$ is much less than the reduction in $\exp(-\Delta G^0/RT)$ from Eq.~(\ref{equ1}). The experiments therefore strongly suggest, in fact, that hairpins lead to a large {\it increase} in the melting rate $k_-$. Since the stable hairpins must first open in order for strands to form a duplex, it is not surprising that they slow down hybridization rates. What is much less obvious is why they so strongly affect melting rates. 

To explore this observation, we use oxDNA~\cite{Ouldridge2011,Ouldridge_thesis,Sulc2012}, a coarse-grained nucleotide-level model of DNA, to simulate and study the systems of Gao {\it et al.}~\cite{Gao2006}. The advantage of simulation is the ability to probe the reactive trajectories in a high level of detail while also capturing the thermodynamics of each system. In particular, we calculate changes in the rate of hybridization and melting due to the effects of hairpins, as well as providing a microscopic understanding of the changes induced by the metastable hairpins.  

The oxDNA model uses pairwise interactions between nucleotides to represent base-pairing, stacking, excluded volume constraints and chain connectivity. The model is discussed in detail elsewhere~\cite{Ouldridge2011,Ouldridge_thesis,Sulc2012},
and has been highly successful at capturing some of the fundamental biophysics
of DNA, including the kinetics of hybridization~\cite{Ouldridge_binding_2013}
and toehold-mediated strand displacement~\cite{srinivas2013}, 
as well as providing insights into nanotechnological 
systems \cite{doye2013coarse}. 

Thermodynamic results presented in this work are obtained using the efficient Virtual-Move Monte Carlo (VMMC) algorithm~\cite{Whitelam2007, whitelam2009role}, augmented with umbrella sampling~\cite{Torrie1977}, to overcome barriers between free-energy minima. Kinetic results are obtained with molecular dynamics using an Andersen-like thermostat~\cite{Russo2009}, which generates diffusive motion of particles beyond a certain (extremely short) timescale. We use forward flux sampling (FFS)~\cite{Allen2005, Allen2009} to determine the rates of hybridization. Before all data collection, we perform lengthy equilibration runs to ensure that the single strands are initialized in thermodynamically representative states. All dynamical simulations were performed at $T_{\rm hyb} =20^\circ$C. Additional simulation details can be found in the supplemental materials.

Using the same strands and terminology of Gao {\it et al.}~\cite{Gao2006}, we consider probe strands P designed to be complementary to target strands T. The pair P$_0$\,T$_0$ is designed to have negligible secondary structure in the single-stranded state, and P$_3$\,T$_3$ and P$_4$\,T$_4$ are designed to have  stable 3- and 4-base pair hairpins  prior to duplex formation. Other than the presence of hairpins, the strands are very similar, having the same length and possessing the same GC-content and similar hybridization free energies relative to an unstructured single strand (as predicted by NuPack~\cite{Dirks2007} and measured with oxDNA). It is therefore reasonable to assume that differences in hybridization rates are primarily due to the presence of hairpins. 

    \begin{figure}
	\begin{center}
	\vspace{0.6 cm}
	\includegraphics[width = 220 pt ]{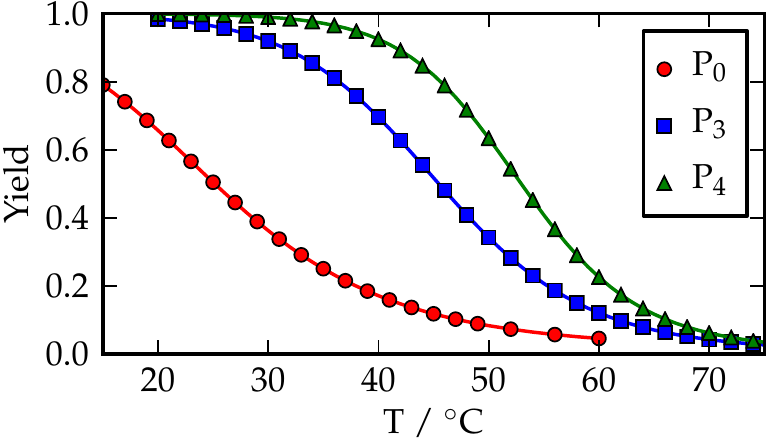}
	\caption{ Equilibrium yields of the hairpins versus temperature for the $P$ strand in each system. Similar results are found for the $T$ strands. }
	\label{yield}
	\end{center}
	\end{figure}

In Fig.~\ref{yield} we plot the hairpin yield curves for the three P strands, obtained using VMMC assisted by umbrella sampling (the results for the T strands are similar). States are designated as hairpins if they possess at least one intra-strand base pair. $\text{P}_0$ and  $\text{T}_0$ have no well-defined strong secondary structure, although a number of transient hairpin configurations contribute to the `hairpin state'. The computed melting temperature of {24}$^\circ$C is only marginally above $20^\circ$C, the temperature used for hybridization measurements, and hairpins are only present  $\sim 60$\% of the time at $20^\circ$C. By contrast, the computed melting temperatures for $\text{P}_3$ and $\text{P}_4$ secondary structure are approximately \ang{46}C and \ang{54}C, respectively. Both systems are dominated by hairpins with approximate yields of 99.8\%  for $\text{P}_4$ and 98.4\% for $\text{P}_3$ at $20^\circ$C. The $\text{P}_3$ and  $\text{T}_3$  strands exhibit variable structure because each strand contains several possible 3-base pair hairpins with varying GC and AT content (consistent with the predictions of NuPack~\cite{Dirks2007}), while $\text{P}_4$ and  $\text{T}_4$ strands are each dominated by a single 4-base pair hairpin containing three GC bonds and one AT bond as designed.  These melting curves are also broadly consistent with those measured by Gao {\it et al.}~\cite{Gao2006}.

    \begin{figure}
    \begin{center}
    \vspace{0.6 cm}
    \includegraphics[width = 220 pt ]{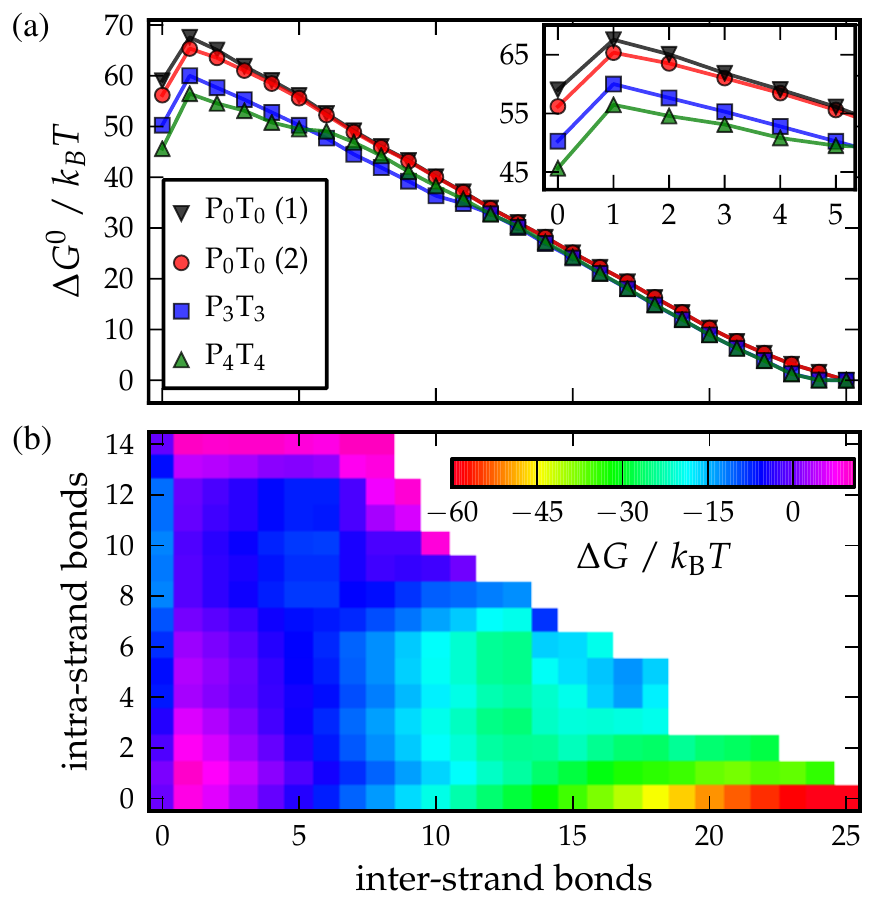}
    \caption{ (Color online) (a) The free-energy profiles for $\text{P}_0 \text{T}_0$, $\text{P}_3 \text{T}_3$ and $\text{P}_4 \text{T}_4$ versus inter-strand bonds between P and T strands. For $\text{P}_0 \text{T}_0$, curve (1) is obtained when hairpin formation is forbidden, and curve (2) in a normal simulation. (b) The free-energy profile for $\text{P}_4 \text{T}_4$ as a function of intra-strand bonds and inter-strand bonds. Both figures have a common $x$-axis. Each system was simulated in a box with a volume of \num{3.96e-23}\SI{}{\cubic\metre} at $T =$ \ang{20}C.}
    \label{free_energy}
    \end{center}
    \end{figure}

In Fig.~\ref{free_energy}(a), we plot free-energy profiles for the three duplex systems as a function of the number of inter-strand bonds, while in Fig.~\ref{free_energy}(b) a two-dimensional (2D) free-energy landscape for the P$_4$T$_4$ system is shown as a function of inter- and intra-strand bonds (2D profiles are shown for the other systems in Fig.~\ref{p3t3_free}). The first figure clearly shows that hairpin formation in each single strand lowers the free energy of the single-stranded state relative to the fully formed duplex, specifically, by approximately 2.2\,$k_{B}T$, 8.6\,$k_{B}T$  and 13.4\,$k_{B}T$ for $\text{P}_0 \text{T}_0$, $\text{P}_3 \text{T}_3$ and $\text{P}_4 \text{T}_4$ respectively. If these differences in single-strand stability were to be manifest only in $k_+$, we would expect hybridization to be slowed by a factor of $\sim600$ for $\text{P}_3 \text{T}_3$ and by a factor of $\sim 7 \times 10^4$ for $\text{P}_4 \text{T}_4$ relative to $\text{P}_0 \text{T}_0$. 

\begin{table}
\renewcommand{\arraystretch}{1.2}
\begin{ruledtabular}
\begin{tabular}{c D{,}{\pm}{-1} D{,}{\pm}{-1} D{,}{\pm}{4.4} }

Duplex 
&  \multicolumn{1}{c}{Experiment $k^{0}_{+}$/$k_{+}$}  
&  \multicolumn{1}{c}{oxDNA $k^{0}_{+}$/$k_{+}$} 
&  \multicolumn{1}{c}{oxDNA $k_{-}$/$k_{-}^{0}$} \\ 
\hline
		$\text{P}_0 \text{T}_0$ & 1 & 1 & 1 \\
        $\text{P}_3 \text{T}_3$ & 1.8,0.2 &  2.1,0.3 & 100,25  \\
        $\text{P}_4 \text{T}_4$ & 6.0,0.3 (f) & 10.6,3.4 & 1934,712 \\
         &  25.0,1.3 (s)  &  & ~ \\ \hline
         & \multicolumn{1}{c}{\shortstack{$k^{0}_{\scriptscriptstyle +}$/$k_{\scriptscriptstyle +}^{\vphantom{0}}$ ($\lambda_0$ $\rightarrow$ $\lambda_1$)}} 
         & \multicolumn{1}{c}{\shortstack{$k^{0}_{\scriptscriptstyle +}$/$k_{\scriptscriptstyle +}^{\vphantom{0}}$ ($\lambda_1$ $\rightarrow$ $\lambda_{2}$)}}
         & \multicolumn{1}{c}{$K_{\text{eq}}^{0}$/$K_{\text{eq}}^{\vphantom{0}}$} \\  \hline
        $\text{P}_0 \text{T}_0$ & 1 & 1 & 1\\
        $\text{P}_3 \text{T}_3$ & 1.85,0.17 &  1.14,0.19 & 204,45\\
        $\text{P}_4 \text{T}_4$ & 2.11,0.11 & 4.93,0.33 & 20539,4096 \\ 
\end{tabular}
\vspace{0.025cm}
\caption{(Top row) Hybridization ($k_{+}$) and melting ($k_{-}$) rate constants for $\text{P}_3 \text{T}_3$ and $\text{P}_4 \text{T}_4$ duplexes, all measured relative to $\text{P}_0 \text{T}_0$, for the experiment~\cite{Gao2006} ($k_{+}$ only) and oxDNA. In the case of $\text{P}_4 \text{T}_4$, Gao {\it et al.} claimed to measure a fast (f) and a slow (s) regime \cite{Gao2006}. (Bottom row) The measured probabilities relative to $\text{P}_0 \text{T}_0$ that a state starting from $\lambda_x$ goes to $\lambda_{x+1}$, where $\lambda_0$, $\lambda_1$, and $\lambda_2$ refer to states which contain 0, 1, and 25 inter-strand bonds. Also listed are the ratios of the calculated equilibrium constants measured relative to $\text{P}_0 \text{T}_0$.}
\end{ruledtabular}
\label{rates_table}
\end{table}

In Table~1 we give the relative hybridization rates for the three systems as
simulated by oxDNA compared to those reported by Gao {\it et
al.}~\cite{Gao2006}. Clearly, the slowdown of hybridization rates is comparable in
simulation and experiment. We note that for $\text{P}_4 \text{T}_4$ Gao {\it et
al.} seemed to observe two regimes with apparently distinct rate constants.
Such behavior might be indicative of a long-lived metastable intermediate
state. We observe metastable intermediates during our simulations but they are
not sufficiently long-lived to result in deviations from simple second-order
kinetics at the strand concentrations used by Gao {\it et al.}  
However, Gao {\it et al.} predict simple second-order behavior at lower 
concentrations with the $\text{P}_3 \text{T}_3$ and $\text{P}_4 \text{T}_4$ duplex formation rates slowed by factors of $\sim 2$ and $\sim 6$ relative to $\text{P}_0 \text{T}_0$. The predictions of oxDNA are quantitatively similar in
this second-order regime.

Possible contributions to the decrease in $k_{+}$ include a reduction in the initial association rate due to fewer bases being accessible for bonding, and also a reduction in the probability that a partially hybridized structure leads to a full duplex. To determine their relative roles, we decomposed the hybridization rate for each system into the rate of initial association, and the probability that an initial association leads to the successful formation of the full duplex. Table~1 shows that the presence of the hairpins slows the rate of association for $\text{P}_3 \text{T}_3$ and $\text{P}_4 \text{T}_4$ by roughly a factor of two in each case when compared to the hairpin-free system. 
Additionally, the probability that the first bond between the two strands leads to a duplex is roughly the same for $\text{P}_0 \text{T}_0$ and $\text{P}_3 \text{T}_3$ systems, while for $\text{P}_4 \text{T}_4$, the initial bond is $\sim$5 times less likely to lead to a full duplex when compared to the hairpins-free case. 

\begin{figure}
    \begin{center}
    \vspace{0.6 cm}
    \includegraphics[width = 220 pt ]{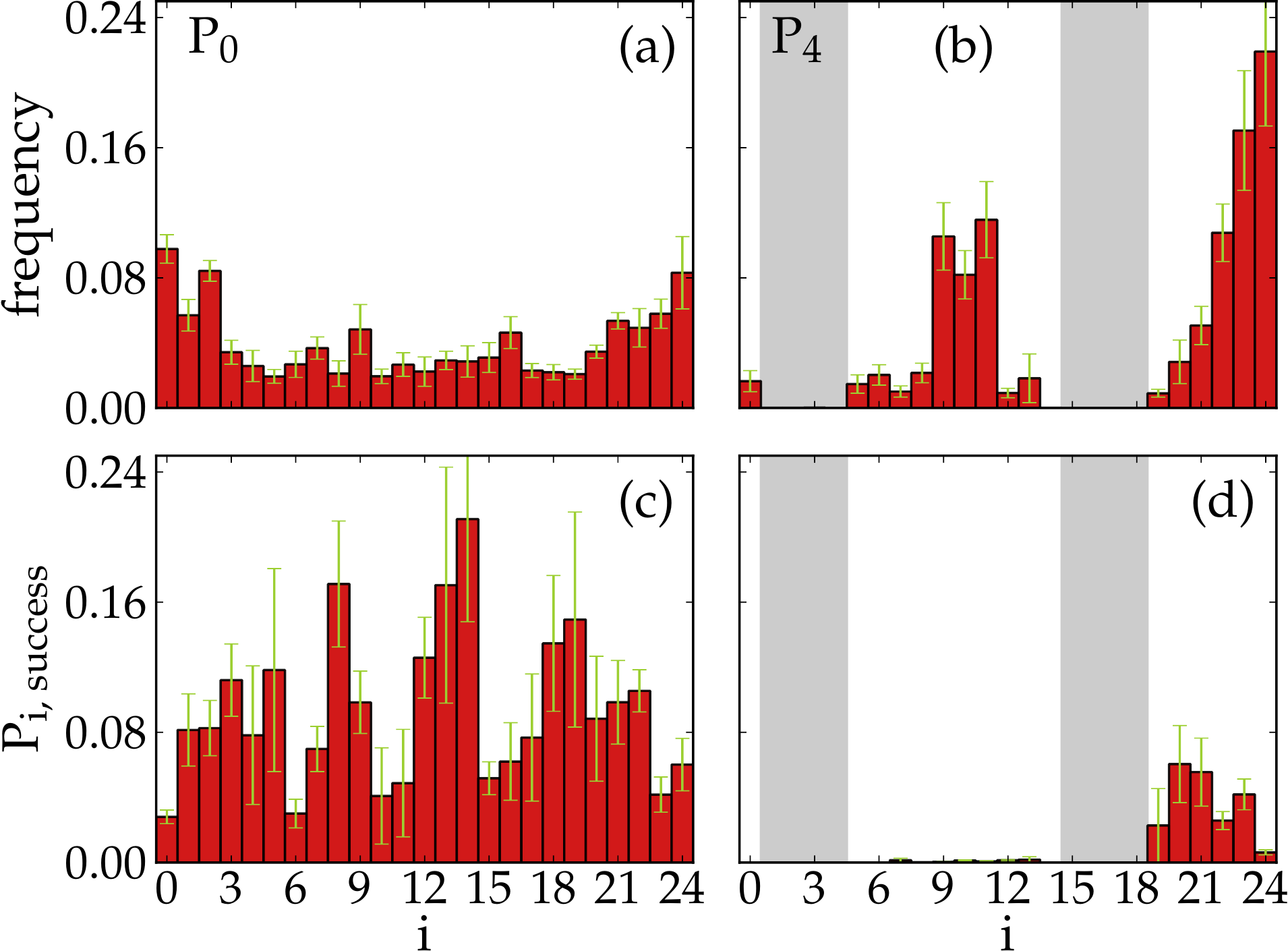}
    \caption{ Top row: Probability of an initial base pair contact between two strands involving base $i$ on the P strand (counting from the $3^\prime$ end), for the (a) $\text{P}_0 \text{T}_0$ and (b) $\text{P}_4 \text{T}_4$ systems. Bottom row: Probability that a contact at position $i$ will subsequently lead to full duplex formation for (c) $\text{P}_0 \text{T}_0$ and (d) $\text{P}_4 \text{T}_4$. The grayed out regions for  $\text{P}_4$ indicate the 4-base pair hairpin stem. Equivalent data for $\text{P}_3 \text{T}_3$ is given in the supplemental material. }
    \label{contact_and_success}
    \end{center}
    \end{figure}

To analyze the initial association events further, in Fig.~\ref{contact_and_success}(a) and (b) we plot the frequency with which initial inter-strand base pairs form at different points on the P strand, for $\text{P}_0 \text{T}_0$ and $\text{P}_4 \text{T}_4$. For $\text{P}_0 \text{T}_0$, there is little systematic variation, except for a bias towards forming base pairs at the strand ends~\cite{Ouldridge_binding_2013}. For $\text{P}_4 \text{T}_4$, however, the bases involved in the hairpin stem never form the initial contacts. Bases that are adjacent to the stem are also harder to access, meaning that initial base pairs tend to form near the apex of the hairpin loop or in a dangling single-stranded tail that is present in both hairpins of the $\text{P}_4 \text{T}_4$ system. 
This reduction in the number of feasible initial binding sites is also evident
from Fig.~\ref{free_energy} where the free-energy barrier separating bound and
unbound states increases by $\sim0.6 k_{\text{B}}T$ and $\sim1.6 k_{\text{B}}T$
for P$_3$T$_3$ and P$_4$T$_4$, respectively, as compared to the same quantity
for P$_0$T$_0$.  

    \begin{figure*}
    \begin{center}
    \vspace{0.6 cm}
    \includegraphics[width = 2\columnwidth ]{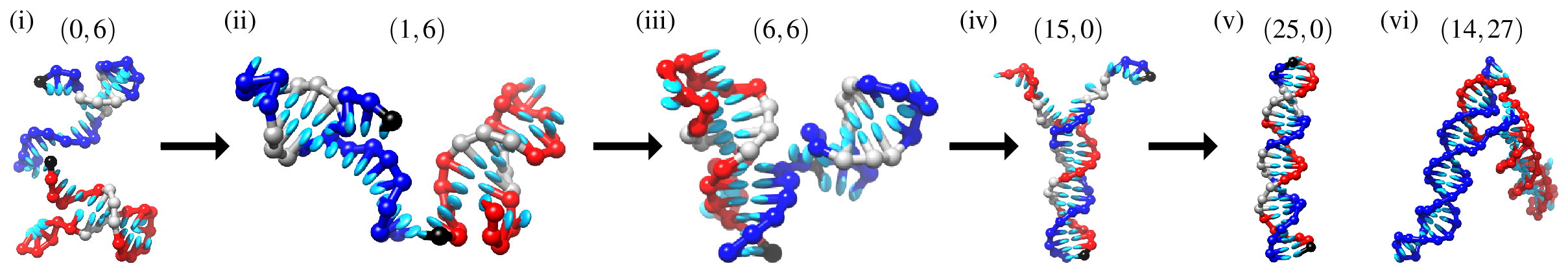}
    \caption{(i-v) Typical duplex formation pathway for $\text{P}_3\text{T}_3$. The black nucleotides indicate the 3$^\prime$ end of the strand. Gray nucleotides indicate the locations of the 3-base pair hairpins present during the association event. (vi) Two long single-stranded hairpins form a four-way junction after fraying base pairs at the end of their stems. The notation (1,6), for example, refers to 1 inter-strand and 6 intra-strand base pairs in the system.}
    \label{P3T3_pathway}
    \end{center}
    \end{figure*}

The overwhelming majority of duplex
formation events, both for $\text{P}_3 \text{T}_3$ and $\text{P}_4
\text{T}_4$, follow a pathway in which the strands associate with intact
hairpins, as is suggested for $\text{P}_4 \text{T}_4$ in
Fig.~\ref{free_energy}(b). A typical pathway for P$_3$T$_3$, illustrated in
Fig.~\ref{P3T3_pathway}, shows that strands retain their 3-base pair hairpins after initially
binding (\ref{P3T3_pathway}(i-ii)). These hairpins are typically still present when the two strands contain significant duplex structure (\ref{P3T3_pathway}(iii)), but must then melt (\ref{P3T3_pathway}(iv)) before proceeding to a full duplex (\ref{P3T3_pathway}(v)). Similarly, in
the P$_4$T$_4$ system, up to 6 base pairs can form between the longer
tails with no loss of the hairpin bonds (Fig.~\ref{p4t4_pathway}). These states, which reside at the local minima at $\sim(5,8)$ in
Fig.~\ref{free_energy}(b), are then blocked by the hairpins from zippering
up further.
For successful hybridization of $\text{P}_3 \text{T}_3$ and
$\text{P}_4 \text{T}_4$, one of the hairpins needs to first melt. To proceed further, either the second hairpin must spontaneously melt before the first reforms, or the first strand could open the other hairpin by displacement. Zippering up of the duplex to completion can then follow. 

The 3-base pair hairpins offer less impediment to successful duplex completion
for two reasons. Firstly, a 3-base pair hairpin melts considerably faster than
a 4-base pair hairpin. Secondly, the $\text{P}_3 \text{T}_3$ system is able to
form ten base pairs before being blocked by the hairpins (Fig.~\ref{p3t3_pathway}(iv)), and once the system
has reached this stage dissociation is very unlikely. By contrast for
$\text{P}_4 \text{T}_4$, even once six base pairs have formed between the
tails, there is still an $\sim80$\% chance of failure (Fig.~\ref{freq_4}(c)). The earlier
loss of hairpin bonds for $\text{P}_4 \text{T}_4$ is evident from
Fig.~\ref{free_energy}(a), where the profile for $\text{P}_4 \text{T}_4$
approaches that for $\text{P}_0 \text{T}_0$ after fewer inter-strand
base pairs. 

Fig.~\ref{contact_and_success}(c) shows that for $\text{P}_0 \text{T}_0$ all
initial binding sites have a non-zero chance of leading to the full duplex (G-C
base pairs are more likely to succeed). By contrast,
Fig.~\ref{contact_and_success}(d) shows that only the bases in the
single-stranded tails of the $\text{P}_4$ and $\text{T}_4$ hairpins have a reasonable
probability of subsequently leading to full duplex formation (and even for
these bases, the success rate is lower than typical of the $\text{P}_0
\text{T}_0$ case). 
Thus, P$_4$T$_4$ strands are unlikely to form a full duplex if the initial
binding occurs between the hairpin loops. 
Although up to 10 base pairs could potentially form between the loops, such
``kissing complexes''
are topologically and geometrically frustrated and are free-energetically
much less stable than the equivalent duplex with the same number of base
pairs~\cite{Romano2012}. They are therefore much more likely to fail than to
succeed. Even if one of the hairpins were to open, displacement of the other
stem from the internal toehold (i.e. the loop) is significantly harder than for
an external toehold~\cite{Green2008}.
These kissing-hairpin states can increase inter-strand bonds without
compromising hairpin structure by also forming base pairs between the tails.
Importantly, the barrier separating these off-pathway intermediate states from
the on-pathway states is not too high (Fig.~\ref{pseudoknot}), and it is not difficult for
intermediates to melt back to on-pathway states.

As the hybridization reactions of these strands follow second-order kinetics to
a very good approximation in oxDNA at low strand concentrations, and we already
know $k_+$ and $\Delta G^0$, we can use Eq.~(\ref{equ1}) to infer $k_{-}$
directly without additional expensive simulations.  The resulting values of $k_-$ for 3- and 4- base pair hairpins are roughly two and three orders of magnitude larger than for the P$_0$T$_0$ system, respectively, as shown in Table 1. Consequently, these hairpins primarily influence the transition kinetics by increasing $k_-$ rather than suppressing $k_+$, confirming our inference from the experimental data. Although we have not actually simulated melting trajectories, the principle of microscopic reversibility applies, and typical melting trajectories will be the reverse of typical formation trajectories. Thus, we conclude that hairpins are not only present after the strands come into contact during assembly, but also prior to strand separation during melting. As one can see from Fig.~\ref{free_energy}(a), once the duplex starts to melt, hairpins can form and stabilize the partially melted states relative to the fully formed duplex, leading to significantly lower free-energy barriers for melting for $\text{P}_3 \text{T}_3$ and $\text{P}_4 \text{T}_4$ than for $\text{P}_0 \text{T}_0$ (by ~6 and 10 $k_{B} T$, respectively). As a consequence, the system has a much higher rate of detachment than it would in the absence of hairpins.

Our results show that hairpins may be less effective as a deliberate design to prevent hybridization
than might have been hoped for, and that small unintended hairpins present only a minor impediment to assembly processes in DNA nanotechnology because so much of the change in $\Delta G^0$ is
absorbed into $k_-$. Single-stranded tails outside the hairpin stem and loop
are particularly problematic in this regard, as they are the prime nucleation
sites for duplex formation. As a result, such tails
should be avoided when designing metastable fuel. However, even carefully
designed hairpins with no tails, long stems, and small loops will primarily increase
$k_-$. Such hairpins are unlikely to form duplexes by first melting their
entire stem. Instead, if a few base pairs fray in each hairpin, the strands can
form a four-way junction as illustrated in Fig.~\ref{P3T3_pathway}(vi). From
this point, intra-molecular base pairs can be exchanged for inter-molecular base
pairs at very low free-energetic cost, and duplex formation can happen through
a process analogous to strand exchange in Holliday Junctions. Considering the
process in reverse, large partially-formed hairpins at four-way junctions
will help to compensate for disrupted inter-strand base pairs, stabilizing
partially melted states and hence enormously increasing $k_-$.

In summary, we have shown numerical evidence that the presence of strong
hairpins in two complementary single strands only marginally affects their
hybridization rates, while it increases melting rates by orders of magnitude.
Hybridization is slightly slowed down both because potential binding sites are
hidden by the secondary structure, and because the hairpins reduce the
probability of initial contacts leading to a full duplex by interfering with
the `zippering' up of the strands. The latter effect is stronger for hairpins
with longer stems and shorter single-stranded tails.  Melting, on the other
hand, is greatly favored because hairpins can form {\em during} the melting
process, thereby stabilizing the partially melted transition states. Our simulations, which provide a mechanistic basis for a general understanding of how secondary structure influences hybridization and melting, are in quantitative agreement with the experiments of Gao {\it et al.}~\cite{Gao2006}, and can be used to guide the design of DNA active devices that use hairpins to modulate hybridization and dissociation rates. 

The authors are grateful to the Engineering and Physical Sciences Research Council.

\bibliography{myDNA.bib}

\begin{thebibliography}{47}%
\makeatletter
\providecommand \@ifxundefined [1]{%
 \@ifx{#1\undefined}
}%
\providecommand \@ifnum [1]{%
 \ifnum #1\expandafter \@firstoftwo
 \else \expandafter \@secondoftwo
 \fi
}%
\providecommand \@ifx [1]{%
 \ifx #1\expandafter \@firstoftwo
 \else \expandafter \@secondoftwo
 \fi
}%
\providecommand \natexlab [1]{#1}%
\providecommand \enquote  [1]{``#1''}%
\providecommand \bibnamefont  [1]{#1}%
\providecommand \bibfnamefont [1]{#1}%
\providecommand \citenamefont [1]{#1}%
\providecommand \href@noop [0]{\@secondoftwo}%
\providecommand \href [0]{\begingroup \@sanitize@url \@href}%
\providecommand \@href[1]{\@@startlink{#1}\@@href}%
\providecommand \@@href[1]{\endgroup#1\@@endlink}%
\providecommand \@sanitize@url [0]{\catcode `\\12\catcode `\$12\catcode
  `\&12\catcode `\#12\catcode `\^12\catcode `\_12\catcode `\%12\relax}%
\providecommand \@@startlink[1]{}%
\providecommand \@@endlink[0]{}%
\providecommand \url  [0]{\begingroup\@sanitize@url \@url }%
\providecommand \@url [1]{\endgroup\@href {#1}{\urlprefix }}%
\providecommand \urlprefix  [0]{URL }%
\providecommand \Eprint [0]{\href }%
\providecommand \doibase [0]{http://dx.doi.org/}%
\providecommand \selectlanguage [0]{\@gobble}%
\providecommand \bibinfo  [0]{\@secondoftwo}%
\providecommand \bibfield  [0]{\@secondoftwo}%
\providecommand \translation [1]{[#1]}%
\providecommand \BibitemOpen [0]{}%
\providecommand \bibitemStop [0]{}%
\providecommand \bibitemNoStop [0]{.\EOS\space}%
\providecommand \EOS [0]{\spacefactor3000\relax}%
\providecommand \BibitemShut  [1]{\csname bibitem#1\endcsname}%
\let\auto@bib@innerbib\@empty
\bibitem [{\citenamefont {Seeman}(1982)}]{Seeman1982}%
  \BibitemOpen
  \bibfield  {author} {\bibinfo {author} {\bibfnamefont {N.~C.}\ \bibnamefont
  {Seeman}},\ }\href@noop {} {\bibfield  {journal} {\bibinfo  {journal} {J.
  Theor. Biol.}\ }\textbf {\bibinfo {volume} {99}},\ \bibinfo {pages} {237}
  (\bibinfo {year} {1982})}\BibitemShut {NoStop}%
\bibitem [{\citenamefont {Rothemund}(2006)}]{Rothemund06}%
  \BibitemOpen
  \bibfield  {author} {\bibinfo {author} {\bibfnamefont {P.~W.~K.}\
  \bibnamefont {Rothemund}},\ }\href@noop {} {\bibfield  {journal} {\bibinfo
  {journal} {Nature}\ }\textbf {\bibinfo {volume} {440}},\ \bibinfo {pages}
  {297} (\bibinfo {year} {2006})}\BibitemShut {NoStop}%
\bibitem [{\citenamefont {Zheng}\ \emph {et~al.}(2009)\citenamefont {Zheng},
  \citenamefont {Birktoft}, \citenamefont {Chen}, \citenamefont {Wang},
  \citenamefont {Sha}, \citenamefont {Constantinou}, \citenamefont {Ginell},
  \citenamefont {Mao},\ and\ \citenamefont {Seeman}}]{Zheng2009}%
  \BibitemOpen
  \bibfield  {author} {\bibinfo {author} {\bibfnamefont {J.}~\bibnamefont
  {Zheng}}, \bibinfo {author} {\bibfnamefont {J.~J.}\ \bibnamefont {Birktoft}},
  \bibinfo {author} {\bibfnamefont {Y.}~\bibnamefont {Chen}}, \bibinfo {author}
  {\bibfnamefont {T.}~\bibnamefont {Wang}}, \bibinfo {author} {\bibfnamefont
  {R.}~\bibnamefont {Sha}}, \bibinfo {author} {\bibfnamefont {P.~E.}\
  \bibnamefont {Constantinou}}, \bibinfo {author} {\bibfnamefont {S.~L.}\
  \bibnamefont {Ginell}}, \bibinfo {author} {\bibfnamefont {C.}~\bibnamefont
  {Mao}}, \ and\ \bibinfo {author} {\bibfnamefont {N.~C.}\ \bibnamefont
  {Seeman}},\ }\href@noop {} {\bibfield  {journal} {\bibinfo  {journal}
  {Nature}\ }\textbf {\bibinfo {volume} {461}},\ \bibinfo {pages} {74}
  (\bibinfo {year} {2009})}\BibitemShut {NoStop}%
\bibitem [{\citenamefont {Dietz}\ \emph {et~al.}(2009)\citenamefont {Dietz},
  \citenamefont {Douglas},\ and\ \citenamefont {Shih}}]{Dietz2009}%
  \BibitemOpen
  \bibfield  {author} {\bibinfo {author} {\bibfnamefont {H.}~\bibnamefont
  {Dietz}}, \bibinfo {author} {\bibfnamefont {S.~M.}\ \bibnamefont {Douglas}},
  \ and\ \bibinfo {author} {\bibfnamefont {W.~M.}\ \bibnamefont {Shih}},\
  }\href@noop {} {\bibfield  {journal} {\bibinfo  {journal} {Science}\ }\textbf
  {\bibinfo {volume} {325}},\ \bibinfo {pages} {725} (\bibinfo {year}
  {2009})}\BibitemShut {NoStop}%
\bibitem [{\citenamefont {Ke}\ \emph {et~al.}(2012)\citenamefont {Ke},
  \citenamefont {Ong}, \citenamefont {Shih},\ and\ \citenamefont
  {Yin}}]{Ke2012}%
  \BibitemOpen
  \bibfield  {author} {\bibinfo {author} {\bibfnamefont {Y.}~\bibnamefont
  {Ke}}, \bibinfo {author} {\bibfnamefont {L.~L.}\ \bibnamefont {Ong}},
  \bibinfo {author} {\bibfnamefont {W.~M.}\ \bibnamefont {Shih}}, \ and\
  \bibinfo {author} {\bibfnamefont {P.}~\bibnamefont {Yin}},\ }\href@noop {}
  {\bibfield  {journal} {\bibinfo  {journal} {Science}\ }\textbf {\bibinfo
  {volume} {338}},\ \bibinfo {pages} {1177} (\bibinfo {year}
  {2012})}\BibitemShut {NoStop}%
\bibitem [{\citenamefont {Yurke}\ \emph {et~al.}(2000)\citenamefont {Yurke},
  \citenamefont {Turberfield}, \citenamefont {Mills}, \citenamefont {Simmel},\
  and\ \citenamefont {Neumann}}]{Yurke2000}%
  \BibitemOpen
  \bibfield  {author} {\bibinfo {author} {\bibfnamefont {B.}~\bibnamefont
  {Yurke}}, \bibinfo {author} {\bibfnamefont {A.~J.}\ \bibnamefont
  {Turberfield}}, \bibinfo {author} {\bibfnamefont {A.~P.}\ \bibnamefont
  {Mills}}, \bibinfo {author} {\bibfnamefont {F.~C.}\ \bibnamefont {Simmel}}, \
  and\ \bibinfo {author} {\bibfnamefont {J.}~\bibnamefont {Neumann}},\
  }\href@noop {} {\bibfield  {journal} {\bibinfo  {journal} {Nature}\ }\textbf
  {\bibinfo {volume} {406}},\ \bibinfo {pages} {605} (\bibinfo {year}
  {2000})}\BibitemShut {NoStop}%
\bibitem [{\citenamefont {Venkataraman}\ \emph {et~al.}(2007)\citenamefont
  {Venkataraman}, \citenamefont {Dirks}, \citenamefont {Rothemund},
  \citenamefont {Winfree},\ and\ \citenamefont {Pierce}}]{Venkataraman2007}%
  \BibitemOpen
  \bibfield  {author} {\bibinfo {author} {\bibfnamefont {S.}~\bibnamefont
  {Venkataraman}}, \bibinfo {author} {\bibfnamefont {R.~M.}\ \bibnamefont
  {Dirks}}, \bibinfo {author} {\bibfnamefont {P.~W.~K.}\ \bibnamefont
  {Rothemund}}, \bibinfo {author} {\bibfnamefont {E.}~\bibnamefont {Winfree}},
  \ and\ \bibinfo {author} {\bibfnamefont {N.~A.}\ \bibnamefont {Pierce}},\
  }\href@noop {} {\bibfield  {journal} {\bibinfo  {journal} {Nat.
  Nanotechnol.}\ }\textbf {\bibinfo {volume} {2}},\ \bibinfo {pages} {490}
  (\bibinfo {year} {2007})}\BibitemShut {NoStop}%
\bibitem [{\citenamefont {Bath}\ \emph {et~al.}(2005)\citenamefont {Bath},
  \citenamefont {Green},\ and\ \citenamefont {Turberfield}}]{Bath2005}%
  \BibitemOpen
  \bibfield  {author} {\bibinfo {author} {\bibfnamefont {J.}~\bibnamefont
  {Bath}}, \bibinfo {author} {\bibfnamefont {S.~J.}\ \bibnamefont {Green}}, \
  and\ \bibinfo {author} {\bibfnamefont {A.~J.}\ \bibnamefont {Turberfield}},\
  }\href@noop {} {\bibfield  {journal} {\bibinfo  {journal} {Angew. Chem. Int.
  Ed.}\ }\textbf {\bibinfo {volume} {117}},\ \bibinfo {pages} {4432} (\bibinfo
  {year} {2005})}\BibitemShut {NoStop}%
\bibitem [{\citenamefont {Bath}\ \emph {et~al.}(2009)\citenamefont {Bath},
  \citenamefont {Green}, \citenamefont {Allan},\ and\ \citenamefont
  {Turberfield}}]{Bath2009}%
  \BibitemOpen
  \bibfield  {author} {\bibinfo {author} {\bibfnamefont {J.}~\bibnamefont
  {Bath}}, \bibinfo {author} {\bibfnamefont {S.~J.}\ \bibnamefont {Green}},
  \bibinfo {author} {\bibfnamefont {K.~E.}\ \bibnamefont {Allan}}, \ and\
  \bibinfo {author} {\bibfnamefont {A.~J.}\ \bibnamefont {Turberfield}},\
  }\href@noop {} {\bibfield  {journal} {\bibinfo  {journal} {Small}\ }\textbf
  {\bibinfo {volume} {5}},\ \bibinfo {pages} {1513} (\bibinfo {year}
  {2009})}\BibitemShut {NoStop}%
\bibitem [{\citenamefont {Omabegho}\ \emph {et~al.}(2009)\citenamefont
  {Omabegho}, \citenamefont {Sha},\ and\ \citenamefont
  {Seeman}}]{Omabegho2009}%
  \BibitemOpen
  \bibfield  {author} {\bibinfo {author} {\bibfnamefont {T.}~\bibnamefont
  {Omabegho}}, \bibinfo {author} {\bibfnamefont {R.}~\bibnamefont {Sha}}, \
  and\ \bibinfo {author} {\bibfnamefont {N.~C.}\ \bibnamefont {Seeman}},\
  }\href@noop {} {\bibfield  {journal} {\bibinfo  {journal} {Science}\ }\textbf
  {\bibinfo {volume} {324}},\ \bibinfo {pages} {67} (\bibinfo {year}
  {2009})}\BibitemShut {NoStop}%
\bibitem [{\citenamefont {Douglas}\ \emph {et~al.}(2012)\citenamefont
  {Douglas}, \citenamefont {Bachelet},\ and\ \citenamefont
  {Church}}]{Douglas2012}%
  \BibitemOpen
  \bibfield  {author} {\bibinfo {author} {\bibfnamefont {S.~M.}\ \bibnamefont
  {Douglas}}, \bibinfo {author} {\bibfnamefont {I.}~\bibnamefont {Bachelet}}, \
  and\ \bibinfo {author} {\bibfnamefont {G.~M.}\ \bibnamefont {Church}},\
  }\href@noop {} {\bibfield  {journal} {\bibinfo  {journal} {Science}\ }\textbf
  {\bibinfo {volume} {335}},\ \bibinfo {pages} {831} (\bibinfo {year}
  {2012})}\BibitemShut {NoStop}%
\bibitem [{\citenamefont {Seelig}\ \emph {et~al.}(2006)\citenamefont {Seelig},
  \citenamefont {Soloveichik}, \citenamefont {Zhang},\ and\ \citenamefont
  {Winfree}}]{Seelig2006}%
  \BibitemOpen
  \bibfield  {author} {\bibinfo {author} {\bibfnamefont {G.}~\bibnamefont
  {Seelig}}, \bibinfo {author} {\bibfnamefont {D.}~\bibnamefont {Soloveichik}},
  \bibinfo {author} {\bibfnamefont {D.~Y.}\ \bibnamefont {Zhang}}, \ and\
  \bibinfo {author} {\bibfnamefont {E.}~\bibnamefont {Winfree}},\ }\href@noop
  {} {\bibfield  {journal} {\bibinfo  {journal} {Science}\ }\textbf {\bibinfo
  {volume} {314}},\ \bibinfo {pages} {1585} (\bibinfo {year}
  {2006})}\BibitemShut {NoStop}%
\bibitem [{\citenamefont {Qian}\ and\ \citenamefont
  {Winfree}(2011)}]{Qian2011}%
  \BibitemOpen
  \bibfield  {author} {\bibinfo {author} {\bibfnamefont {L.}~\bibnamefont
  {Qian}}\ and\ \bibinfo {author} {\bibfnamefont {E.}~\bibnamefont {Winfree}},\
  }\href@noop {} {\bibfield  {journal} {\bibinfo  {journal} {Science}\ }\textbf
  {\bibinfo {volume} {332}},\ \bibinfo {pages} {1196} (\bibinfo {year}
  {2011})}\BibitemShut {NoStop}%
\bibitem [{\citenamefont {Chen}\ \emph {et~al.}(2013)\citenamefont {Chen},
  \citenamefont {Dalchau}, \citenamefont {Srinivas}, \citenamefont {Phillips},
  \citenamefont {Cardelli}, \citenamefont {Solveichik},\ and\ \citenamefont
  {Seelig}}]{Chen2013}%
  \BibitemOpen
  \bibfield  {author} {\bibinfo {author} {\bibfnamefont {Y.-J.}\ \bibnamefont
  {Chen}}, \bibinfo {author} {\bibfnamefont {N.}~\bibnamefont {Dalchau}},
  \bibinfo {author} {\bibfnamefont {N.}~\bibnamefont {Srinivas}}, \bibinfo
  {author} {\bibfnamefont {A.}~\bibnamefont {Phillips}}, \bibinfo {author}
  {\bibfnamefont {L.}~\bibnamefont {Cardelli}}, \bibinfo {author}
  {\bibfnamefont {D.}~\bibnamefont {Solveichik}}, \ and\ \bibinfo {author}
  {\bibfnamefont {G.}~\bibnamefont {Seelig}},\ }\href@noop {} {\bibfield
  {journal} {\bibinfo  {journal} {Nat. Nanotechnol.}\ }\textbf {\bibinfo
  {volume} {8}},\ \bibinfo {pages} {755} (\bibinfo {year} {2013})}\BibitemShut
  {NoStop}%
\bibitem [{\citenamefont {Gao}\ \emph {et~al.}(2006)\citenamefont {Gao},
  \citenamefont {Wolf},\ and\ \citenamefont {Georgiadis}}]{Gao2006}%
  \BibitemOpen
  \bibfield  {author} {\bibinfo {author} {\bibfnamefont {Y.}~\bibnamefont
  {Gao}}, \bibinfo {author} {\bibfnamefont {L.~K.}\ \bibnamefont {Wolf}}, \
  and\ \bibinfo {author} {\bibfnamefont {R.~M.}\ \bibnamefont {Georgiadis}},\
  }\href@noop {} {\bibfield  {journal} {\bibinfo  {journal} {Nucl. Acids Res.}\
  }\textbf {\bibinfo {volume} {34}},\ \bibinfo {pages} {3370} (\bibinfo {year}
  {2006})}\BibitemShut {NoStop}%
\bibitem [{\citenamefont {Romano}\ \emph {et~al.}(2013)\citenamefont {Romano},
  \citenamefont {Chakraborty}, \citenamefont {Doye}, \citenamefont
  {Ouldridge},\ and\ \citenamefont {Louis}}]{Romano_overstretch_2013}%
  \BibitemOpen
  \bibfield  {author} {\bibinfo {author} {\bibfnamefont {F.}~\bibnamefont
  {Romano}}, \bibinfo {author} {\bibfnamefont {D.}~\bibnamefont {Chakraborty}},
  \bibinfo {author} {\bibfnamefont {J.~P.~K.}\ \bibnamefont {Doye}}, \bibinfo
  {author} {\bibfnamefont {T.~E.}\ \bibnamefont {Ouldridge}}, \ and\ \bibinfo
  {author} {\bibfnamefont {A.~A.}\ \bibnamefont {Louis}},\ }\href@noop {}
  {\bibfield  {journal} {\bibinfo  {journal} {J. Chem. Phys.}\ }\textbf
  {\bibinfo {volume} {138}},\ \bibinfo {pages} {085101} (\bibinfo {year}
  {2013})}\BibitemShut {NoStop}%
\bibitem [{\citenamefont {Turberfield}\ \emph {et~al.}(2003)\citenamefont
  {Turberfield}, \citenamefont {Mitchell}, \citenamefont {Yurke}, \citenamefont
  {Mills}, \citenamefont {Blakey},\ and\ \citenamefont
  {Simmel}}]{Turberfield2003}%
  \BibitemOpen
  \bibfield  {author} {\bibinfo {author} {\bibfnamefont {A.~J.}\ \bibnamefont
  {Turberfield}}, \bibinfo {author} {\bibfnamefont {J.~C.}\ \bibnamefont
  {Mitchell}}, \bibinfo {author} {\bibfnamefont {B.}~\bibnamefont {Yurke}},
  \bibinfo {author} {\bibfnamefont {A.~P.}\ \bibnamefont {Mills}}, \bibinfo
  {author} {\bibfnamefont {M.~I.}\ \bibnamefont {Blakey}}, \ and\ \bibinfo
  {author} {\bibfnamefont {F.~C.}\ \bibnamefont {Simmel}},\ }\href@noop {}
  {\bibfield  {journal} {\bibinfo  {journal} {Phys. Rev. Lett.}\ }\textbf
  {\bibinfo {volume} {90}},\ \bibinfo {pages} {118102} (\bibinfo {year}
  {2003})}\BibitemShut {NoStop}%
\bibitem [{\citenamefont {Green}\ \emph {et~al.}(2008)\citenamefont {Green},
  \citenamefont {Bath},\ and\ \citenamefont {Turberfield}}]{Green2008}%
  \BibitemOpen
  \bibfield  {author} {\bibinfo {author} {\bibfnamefont {S.~J.}\ \bibnamefont
  {Green}}, \bibinfo {author} {\bibfnamefont {J.}~\bibnamefont {Bath}}, \ and\
  \bibinfo {author} {\bibfnamefont {A.~J.}\ \bibnamefont {Turberfield}},\
  }\href@noop {} {\bibfield  {journal} {\bibinfo  {journal} {Phys. Rev. Lett.}\
  }\textbf {\bibinfo {volume} {101}},\ \bibinfo {pages} {238101} (\bibinfo
  {year} {2008})}\BibitemShut {NoStop}%
\bibitem [{\citenamefont {Tomov}\ \emph {et~al.}(2013)\citenamefont {Tomov},
  \citenamefont {Tsukanov}, \citenamefont {Liber}, \citenamefont {Masoud},
  \citenamefont {Plavner},\ and\ \citenamefont {Nir}}]{Tomov2013}%
  \BibitemOpen
  \bibfield  {author} {\bibinfo {author} {\bibfnamefont {T.~E.}\ \bibnamefont
  {Tomov}}, \bibinfo {author} {\bibfnamefont {R.}~\bibnamefont {Tsukanov}},
  \bibinfo {author} {\bibfnamefont {M.}~\bibnamefont {Liber}}, \bibinfo
  {author} {\bibfnamefont {R.}~\bibnamefont {Masoud}}, \bibinfo {author}
  {\bibfnamefont {N.}~\bibnamefont {Plavner}}, \ and\ \bibinfo {author}
  {\bibfnamefont {E.}~\bibnamefont {Nir}},\ }\href@noop {} {\bibfield
  {journal} {\bibinfo  {journal} {J. Am. Chem. Soc.}\ }\textbf {\bibinfo
  {volume} {135}},\ \bibinfo {pages} {11935} (\bibinfo {year}
  {2013})}\BibitemShut {NoStop}%
\bibitem [{\citenamefont {Ouldridge}\ \emph {et~al.}(2011)\citenamefont
  {Ouldridge}, \citenamefont {Louis},\ and\ \citenamefont
  {Doye}}]{Ouldridge2011}%
  \BibitemOpen
  \bibfield  {author} {\bibinfo {author} {\bibfnamefont {T.~E.}\ \bibnamefont
  {Ouldridge}}, \bibinfo {author} {\bibfnamefont {A.~A.}\ \bibnamefont
  {Louis}}, \ and\ \bibinfo {author} {\bibfnamefont {J.~P.~K.}\ \bibnamefont
  {Doye}},\ }\href@noop {} {\bibfield  {journal} {\bibinfo  {journal} {J. Chem.
  Phys.}\ }\textbf {\bibinfo {volume} {134}},\ \bibinfo {pages} {085101}
  (\bibinfo {year} {2011})}\BibitemShut {NoStop}%
\bibitem [{\citenamefont {Ouldridge}(2012{\natexlab{a}})}]{Ouldridge_thesis}%
  \BibitemOpen
  \bibfield  {author} {\bibinfo {author} {\bibfnamefont {T.~E.}\ \bibnamefont
  {Ouldridge}},\ }\emph {\bibinfo {title} {Coarse-grained modelling of {DNA}
  and {DNA} nanotechnology}},\ \href@noop {} {Ph.D. thesis},\ \bibinfo
  {school} {University of Oxford} (\bibinfo {year} {2011 [Published as a book
  by Springer, Heidelberg, 2012]}{\natexlab{a}})\BibitemShut {NoStop}%
\bibitem [{\citenamefont {{\v S}ulc}\ \emph {et~al.}(2012)\citenamefont {{\v
  S}ulc}, \citenamefont {Romano}, \citenamefont {Ouldridge}, \citenamefont
  {Rovigatti}, \citenamefont {Doye},\ and\ \citenamefont {Louis}}]{Sulc2012}%
  \BibitemOpen
  \bibfield  {author} {\bibinfo {author} {\bibfnamefont {P.}~\bibnamefont {{\v
  S}ulc}}, \bibinfo {author} {\bibfnamefont {F.}~\bibnamefont {Romano}},
  \bibinfo {author} {\bibfnamefont {T.~E.}\ \bibnamefont {Ouldridge}}, \bibinfo
  {author} {\bibfnamefont {L.}~\bibnamefont {Rovigatti}}, \bibinfo {author}
  {\bibfnamefont {J.~P.~K.}\ \bibnamefont {Doye}}, \ and\ \bibinfo {author}
  {\bibfnamefont {A.~A.}\ \bibnamefont {Louis}},\ }\href@noop {} {\bibfield
  {journal} {\bibinfo  {journal} {J. Chem. Phys.}\ }\textbf {\bibinfo {volume}
  {137}},\ \bibinfo {pages} {135101} (\bibinfo {year} {2012})}\BibitemShut
  {NoStop}%
\bibitem [{\citenamefont {Ouldridge}\ \emph
  {et~al.}(2013{\natexlab{a}})\citenamefont {Ouldridge}, \citenamefont
  {\v{S}ulc}, \citenamefont {Romano}, \citenamefont {Doye},\ and\ \citenamefont
  {Louis}}]{Ouldridge_binding_2013}%
  \BibitemOpen
  \bibfield  {author} {\bibinfo {author} {\bibfnamefont {T.~E.}\ \bibnamefont
  {Ouldridge}}, \bibinfo {author} {\bibfnamefont {P.}~\bibnamefont {\v{S}ulc}},
  \bibinfo {author} {\bibfnamefont {F.}~\bibnamefont {Romano}}, \bibinfo
  {author} {\bibfnamefont {J.~P.~K.}\ \bibnamefont {Doye}}, \ and\ \bibinfo
  {author} {\bibfnamefont {A.~A.}\ \bibnamefont {Louis}},\ }\href@noop {}
  {\bibfield  {journal} {\bibinfo  {journal} {Nucl. Acids Res.}\ }\textbf
  {\bibinfo {volume} {41}},\ \bibinfo {pages} {8886} (\bibinfo {year}
  {2013}{\natexlab{a}})}\BibitemShut {NoStop}%
\bibitem [{\citenamefont {Srinivas}\ \emph {et~al.}(2013)\citenamefont
  {Srinivas}, \citenamefont {Ouldridge}, \citenamefont {\v{S}ulc},
  \citenamefont {Schaeffer}, \citenamefont {Yurke}, \citenamefont {Louis},
  \citenamefont {Doye},\ and\ \citenamefont {Winfree}}]{srinivas2013}%
  \BibitemOpen
  \bibfield  {author} {\bibinfo {author} {\bibfnamefont {N.}~\bibnamefont
  {Srinivas}}, \bibinfo {author} {\bibfnamefont {T.~E.}\ \bibnamefont
  {Ouldridge}}, \bibinfo {author} {\bibfnamefont {P.}~\bibnamefont {\v{S}ulc}},
  \bibinfo {author} {\bibfnamefont {J.}~\bibnamefont {Schaeffer}}, \bibinfo
  {author} {\bibfnamefont {B.}~\bibnamefont {Yurke}}, \bibinfo {author}
  {\bibfnamefont {A.~A.}\ \bibnamefont {Louis}}, \bibinfo {author}
  {\bibfnamefont {J.~P.~K.}\ \bibnamefont {Doye}}, \ and\ \bibinfo {author}
  {\bibfnamefont {E.}~\bibnamefont {Winfree}},\ }\href@noop {} {\bibfield
  {journal} {\bibinfo  {journal} {Nucl. Acids Res.}\ }\textbf {\bibinfo
  {volume} {41}},\ \bibinfo {pages} {10641} (\bibinfo {year}
  {2013})}\BibitemShut {NoStop}%
\bibitem [{\citenamefont {Doye}\ \emph {et~al.}(2013)\citenamefont {Doye},
  \citenamefont {Ouldridge}, \citenamefont {Louis}, \citenamefont {Romano},
  \citenamefont {{\v{S}}ulc}, \citenamefont {Matek}, \citenamefont {Snodin},
  \citenamefont {Rovigatti}, \citenamefont {Schreck}, \citenamefont {Harrison}
  \emph {et~al.}}]{doye2013coarse}%
  \BibitemOpen
  \bibfield  {author} {\bibinfo {author} {\bibfnamefont {J.~P.~K.}\
  \bibnamefont {Doye}}, \bibinfo {author} {\bibfnamefont {T.~E.}\ \bibnamefont
  {Ouldridge}}, \bibinfo {author} {\bibfnamefont {A.~A.}\ \bibnamefont
  {Louis}}, \bibinfo {author} {\bibfnamefont {F.}~\bibnamefont {Romano}},
  \bibinfo {author} {\bibfnamefont {P.}~\bibnamefont {{\v{S}}ulc}}, \bibinfo
  {author} {\bibfnamefont {C.}~\bibnamefont {Matek}}, \bibinfo {author}
  {\bibfnamefont {B.~E.}\ \bibnamefont {Snodin}}, \bibinfo {author}
  {\bibfnamefont {L.}~\bibnamefont {Rovigatti}}, \bibinfo {author}
  {\bibfnamefont {J.~S.}\ \bibnamefont {Schreck}}, \bibinfo {author}
  {\bibfnamefont {R.~M.}\ \bibnamefont {Harrison}},  \emph {et~al.},\
  }\href@noop {} {\bibfield  {journal} {\bibinfo  {journal} {Phys. Chem. Chem.
  Phys.}\ }\textbf {\bibinfo {volume} {15}},\ \bibinfo {pages} {20395}
  (\bibinfo {year} {2013})}\BibitemShut {NoStop}%
\bibitem [{\citenamefont {Whitelam}\ and\ \citenamefont
  {Geissler}(2007)}]{Whitelam2007}%
  \BibitemOpen
  \bibfield  {author} {\bibinfo {author} {\bibfnamefont {S.}~\bibnamefont
  {Whitelam}}\ and\ \bibinfo {author} {\bibfnamefont {P.~L.}\ \bibnamefont
  {Geissler}},\ }\href@noop {} {\bibfield  {journal} {\bibinfo  {journal} {J.
  Chem. Phys.}\ }\textbf {\bibinfo {volume} {127}},\ \bibinfo {pages} {154101}
  (\bibinfo {year} {2007})}\BibitemShut {NoStop}%
\bibitem [{\citenamefont {Whitelam}\ \emph {et~al.}(2009)\citenamefont
  {Whitelam}, \citenamefont {Feng}, \citenamefont {Hagan},\ and\ \citenamefont
  {Geissler}}]{whitelam2009role}%
  \BibitemOpen
  \bibfield  {author} {\bibinfo {author} {\bibfnamefont {S.}~\bibnamefont
  {Whitelam}}, \bibinfo {author} {\bibfnamefont {E.~H.}\ \bibnamefont {Feng}},
  \bibinfo {author} {\bibfnamefont {M.~F.}\ \bibnamefont {Hagan}}, \ and\
  \bibinfo {author} {\bibfnamefont {P.~L.}\ \bibnamefont {Geissler}},\
  }\href@noop {} {\bibfield  {journal} {\bibinfo  {journal} {Soft Matter}\
  }\textbf {\bibinfo {volume} {5}},\ \bibinfo {pages} {1251} (\bibinfo {year}
  {2009})}\BibitemShut {NoStop}%
\bibitem [{\citenamefont {Torrie}\ and\ \citenamefont
  {Valleau}(1977{\natexlab{a}})}]{Torrie1977}%
  \BibitemOpen
  \bibfield  {author} {\bibinfo {author} {\bibfnamefont {G.~M.}\ \bibnamefont
  {Torrie}}\ and\ \bibinfo {author} {\bibfnamefont {J.~P.}\ \bibnamefont
  {Valleau}},\ }\href@noop {} {\bibfield  {journal} {\bibinfo  {journal} {J.
  Comp. Phys.}\ }\textbf {\bibinfo {volume} {23}},\ \bibinfo {pages} {187}
  (\bibinfo {year} {1977}{\natexlab{a}})}\BibitemShut {NoStop}%
\bibitem [{\citenamefont {Russo}\ \emph
  {et~al.}(2009{\natexlab{a}})\citenamefont {Russo}, \citenamefont
  {Tartaglia},\ and\ \citenamefont {Sciortino}}]{Russo2009}%
  \BibitemOpen
  \bibfield  {author} {\bibinfo {author} {\bibfnamefont {J.}~\bibnamefont
  {Russo}}, \bibinfo {author} {\bibfnamefont {P.}~\bibnamefont {Tartaglia}}, \
  and\ \bibinfo {author} {\bibfnamefont {F.}~\bibnamefont {Sciortino}},\
  }\href@noop {} {\bibfield  {journal} {\bibinfo  {journal} {J, Chem. Phys.}\
  }\textbf {\bibinfo {volume} {131}},\ \bibinfo {pages} {014504} (\bibinfo
  {year} {2009}{\natexlab{a}})}\BibitemShut {NoStop}%
\bibitem [{\citenamefont {Allen}\ \emph {et~al.}(2005)\citenamefont {Allen},
  \citenamefont {Warren},\ and\ \citenamefont {{ten Wolde}}}]{Allen2005}%
  \BibitemOpen
  \bibfield  {author} {\bibinfo {author} {\bibfnamefont {R.~J.}\ \bibnamefont
  {Allen}}, \bibinfo {author} {\bibfnamefont {P.~B.}\ \bibnamefont {Warren}}, \
  and\ \bibinfo {author} {\bibfnamefont {P.~R.}\ \bibnamefont {{ten Wolde}}},\
  }\href@noop {} {\bibfield  {journal} {\bibinfo  {journal} {Phys. Rev. Lett.}\
  }\textbf {\bibinfo {volume} {94}},\ \bibinfo {pages} {018104} (\bibinfo
  {year} {2005})}\BibitemShut {NoStop}%
\bibitem [{\citenamefont {Allen}\ \emph {et~al.}(2009)\citenamefont {Allen},
  \citenamefont {Valeriani},\ and\ \citenamefont {{ten Wolde}}}]{Allen2009}%
  \BibitemOpen
  \bibfield  {author} {\bibinfo {author} {\bibfnamefont {R.~J.}\ \bibnamefont
  {Allen}}, \bibinfo {author} {\bibfnamefont {C.}~\bibnamefont {Valeriani}}, \
  and\ \bibinfo {author} {\bibfnamefont {P.~R.}\ \bibnamefont {{ten Wolde}}},\
  }\href@noop {} {\bibfield  {journal} {\bibinfo  {journal} {J. Phys.: Condens.
  Matter}\ }\textbf {\bibinfo {volume} {21}},\ \bibinfo {pages} {463102}
  (\bibinfo {year} {2009})}\BibitemShut {NoStop}%
\bibitem [{\citenamefont {Dirks}\ \emph {et~al.}(2007)\citenamefont {Dirks},
  \citenamefont {Bois}, \citenamefont {Schaeffer}, \citenamefont {Winfree},\
  and\ \citenamefont {Pierce}}]{Dirks2007}%
  \BibitemOpen
  \bibfield  {author} {\bibinfo {author} {\bibfnamefont {R.~M.}\ \bibnamefont
  {Dirks}}, \bibinfo {author} {\bibfnamefont {J.~S.}\ \bibnamefont {Bois}},
  \bibinfo {author} {\bibfnamefont {J.~M.}\ \bibnamefont {Schaeffer}}, \bibinfo
  {author} {\bibfnamefont {E.}~\bibnamefont {Winfree}}, \ and\ \bibinfo
  {author} {\bibfnamefont {N.~A.}\ \bibnamefont {Pierce}},\ }\href@noop {}
  {\bibfield  {journal} {\bibinfo  {journal} {SIAM Rev.}\ }\textbf {\bibinfo
  {volume} {29}},\ \bibinfo {pages} {65} (\bibinfo {year} {2007})}\BibitemShut
  {NoStop}%
\bibitem [{\citenamefont {Romano}\ \emph {et~al.}(2012)\citenamefont {Romano},
  \citenamefont {Hudson}, \citenamefont {Doye}, \citenamefont {Ouldridge},\
  and\ \citenamefont {Louis}}]{Romano2012}%
  \BibitemOpen
  \bibfield  {author} {\bibinfo {author} {\bibfnamefont {F.}~\bibnamefont
  {Romano}}, \bibinfo {author} {\bibfnamefont {A.}~\bibnamefont {Hudson}},
  \bibinfo {author} {\bibfnamefont {J.~P.~K.}\ \bibnamefont {Doye}}, \bibinfo
  {author} {\bibfnamefont {T.~E.}\ \bibnamefont {Ouldridge}}, \ and\ \bibinfo
  {author} {\bibfnamefont {A.~A.}\ \bibnamefont {Louis}},\ }\href@noop {}
  {\bibfield  {journal} {\bibinfo  {journal} {J. Chem. Phys.}\ }\textbf
  {\bibinfo {volume} {136}},\ \bibinfo {pages} {215102} (\bibinfo {year}
  {2012})}\BibitemShut {NoStop}%
\bibitem [{\citenamefont {Ouldridge}\ \emph
  {et~al.}(2013{\natexlab{b}})\citenamefont {Ouldridge}, \citenamefont {Hoare},
  \citenamefont {Louis}, \citenamefont {Doye}, \citenamefont {Bath},\ and\
  \citenamefont {Turberfield}}]{Ouldridge_walker_2013}%
  \BibitemOpen
  \bibfield  {author} {\bibinfo {author} {\bibfnamefont {T.~E.}\ \bibnamefont
  {Ouldridge}}, \bibinfo {author} {\bibfnamefont {R.~L.}\ \bibnamefont
  {Hoare}}, \bibinfo {author} {\bibfnamefont {A.~A.}\ \bibnamefont {Louis}},
  \bibinfo {author} {\bibfnamefont {J.~P.~K.}\ \bibnamefont {Doye}}, \bibinfo
  {author} {\bibfnamefont {J.}~\bibnamefont {Bath}}, \ and\ \bibinfo {author}
  {\bibfnamefont {A.~J.}\ \bibnamefont {Turberfield}},\ }\href@noop {}
  {\bibfield  {journal} {\bibinfo  {journal} {ACS Nano}\ }\textbf {\bibinfo
  {volume} {7}},\ \bibinfo {pages} {2479} (\bibinfo {year}
  {2013}{\natexlab{b}})}\BibitemShut {NoStop}%
\bibitem [{\citenamefont {SantaLucia}(1998)}]{SantaLucia1998}%
  \BibitemOpen
  \bibfield  {author} {\bibinfo {author} {\bibfnamefont {J.}~\bibnamefont
  {SantaLucia}, \bibfnamefont {Jr.}},\ }\href@noop {} {\bibfield  {journal}
  {\bibinfo  {journal} {Proc. Natl. Acad. Sci. U.S.A}\ }\textbf {\bibinfo
  {volume} {17}},\ \bibinfo {pages} {1460} (\bibinfo {year}
  {1998})}\BibitemShut {NoStop}%
\bibitem [{\citenamefont {Yakovchuk}\ \emph {et~al.}(2006)\citenamefont
  {Yakovchuk}, \citenamefont {Protozanova},\ and\ \citenamefont
  {{Frank-Kamenetskii}}}]{Yakovchuk2006}%
  \BibitemOpen
  \bibfield  {author} {\bibinfo {author} {\bibfnamefont {P.}~\bibnamefont
  {Yakovchuk}}, \bibinfo {author} {\bibfnamefont {E.}~\bibnamefont
  {Protozanova}}, \ and\ \bibinfo {author} {\bibfnamefont {M.~D.}\ \bibnamefont
  {{Frank-Kamenetskii}}},\ }\href@noop {} {\bibfield  {journal} {\bibinfo
  {journal} {Nucl. Acids Res.}\ }\textbf {\bibinfo {volume} {34}},\ \bibinfo
  {pages} {564} (\bibinfo {year} {2006})}\BibitemShut {NoStop}%
\bibitem [{\citenamefont {Frenkel}\ and\ \citenamefont
  {Smit}(2001)}]{Frenkel2001}%
  \BibitemOpen
  \bibfield  {author} {\bibinfo {author} {\bibfnamefont {D.}~\bibnamefont
  {Frenkel}}\ and\ \bibinfo {author} {\bibfnamefont {B.}~\bibnamefont {Smit}},\
  }\href@noop {} {\emph {\bibinfo {title} {Understanding Molecular
  Simulation}}}\ (\bibinfo  {publisher} {Academic Press Inc. London},\ \bibinfo
  {year} {2001})\BibitemShut {NoStop}%
\bibitem [{\citenamefont {Torrie}\ and\ \citenamefont
  {Valleau}(1977{\natexlab{b}})}]{torrie1977nonphysical}%
  \BibitemOpen
  \bibfield  {author} {\bibinfo {author} {\bibfnamefont {G.~M.}\ \bibnamefont
  {Torrie}}\ and\ \bibinfo {author} {\bibfnamefont {J.~P.}\ \bibnamefont
  {Valleau}},\ }\href@noop {} {\bibfield  {journal} {\bibinfo  {journal} {J.
  Comp. Phys.}\ }\textbf {\bibinfo {volume} {23}},\ \bibinfo {pages} {187}
  (\bibinfo {year} {1977}{\natexlab{b}})}\BibitemShut {NoStop}%
\bibitem [{\citenamefont {Ferrenberg}\ and\ \citenamefont
  {Swendsen}(1988)}]{Ferrenberg1988}%
  \BibitemOpen
  \bibfield  {author} {\bibinfo {author} {\bibfnamefont {A.~M.}\ \bibnamefont
  {Ferrenberg}}\ and\ \bibinfo {author} {\bibfnamefont {R.~H.}\ \bibnamefont
  {Swendsen}},\ }\href@noop {} {\bibfield  {journal} {\bibinfo  {journal}
  {Phys. Rev. Lett.}\ }\textbf {\bibinfo {volume} {61}},\ \bibinfo {pages}
  {2635} (\bibinfo {year} {1988})}\BibitemShut {NoStop}%
\bibitem [{\citenamefont {Russo}\ \emph
  {et~al.}(2009{\natexlab{b}})\citenamefont {Russo}, \citenamefont
  {Tartaglia},\ and\ \citenamefont {Sciortino}}]{russo2009reversible}%
  \BibitemOpen
  \bibfield  {author} {\bibinfo {author} {\bibfnamefont {J.}~\bibnamefont
  {Russo}}, \bibinfo {author} {\bibfnamefont {P.}~\bibnamefont {Tartaglia}}, \
  and\ \bibinfo {author} {\bibfnamefont {F.}~\bibnamefont {Sciortino}},\
  }\href@noop {} {\bibfield  {journal} {\bibinfo  {journal} {J. Chem. Phys.}\
  }\textbf {\bibinfo {volume} {131}},\ \bibinfo {pages} {014504} (\bibinfo
  {year} {2009}{\natexlab{b}})}\BibitemShut {NoStop}%
\bibitem [{\citenamefont {Verlet}(1967)}]{verlet1967computer}%
  \BibitemOpen
  \bibfield  {author} {\bibinfo {author} {\bibfnamefont {L.}~\bibnamefont
  {Verlet}},\ }\href@noop {} {\bibfield  {journal} {\bibinfo  {journal} {Phys.
  Rev.}\ }\textbf {\bibinfo {volume} {159}},\ \bibinfo {pages} {98} (\bibinfo
  {year} {1967})}\BibitemShut {NoStop}%
\bibitem [{\citenamefont {Lapham}\ \emph {et~al.}(1997)\citenamefont {Lapham},
  \citenamefont {Rife}, \citenamefont {Moore},\ and\ \citenamefont
  {Crothers}}]{Lapham1997}%
  \BibitemOpen
  \bibfield  {author} {\bibinfo {author} {\bibfnamefont {J.}~\bibnamefont
  {Lapham}}, \bibinfo {author} {\bibfnamefont {J.~P.}\ \bibnamefont {Rife}},
  \bibinfo {author} {\bibfnamefont {P.~B.}\ \bibnamefont {Moore}}, \ and\
  \bibinfo {author} {\bibfnamefont {D.~M.}\ \bibnamefont {Crothers}},\
  }\href@noop {} {\bibfield  {journal} {\bibinfo  {journal} {J. Biomol. NMR}\
  }\textbf {\bibinfo {volume} {10}},\ \bibinfo {pages} {252} (\bibinfo {year}
  {1997})}\BibitemShut {NoStop}%
\bibitem [{\citenamefont {Murtola}\ \emph {et~al.}(2009)\citenamefont
  {Murtola}, \citenamefont {Bunkwer}, \citenamefont {Vattulainen},\ and\
  \citenamefont {Deserno}}]{Murtola2009}%
  \BibitemOpen
  \bibfield  {author} {\bibinfo {author} {\bibfnamefont {T.}~\bibnamefont
  {Murtola}}, \bibinfo {author} {\bibfnamefont {A.}~\bibnamefont {Bunkwer}},
  \bibinfo {author} {\bibfnamefont {I.}~\bibnamefont {Vattulainen}}, \ and\
  \bibinfo {author} {\bibfnamefont {M.}~\bibnamefont {Deserno}},\ }\href@noop
  {} {\bibfield  {journal} {\bibinfo  {journal} {Phys. Chem. Chem. Phys.}\
  }\textbf {\bibinfo {volume} {11}},\ \bibinfo {pages} {1869} (\bibinfo {year}
  {2009})}\BibitemShut {NoStop}%
\bibitem [{\citenamefont {Ouldridge}\ \emph {et~al.}(2010)\citenamefont
  {Ouldridge}, \citenamefont {Louis},\ and\ \citenamefont
  {Doye}}]{Ouldridge_bulk_2010}%
  \BibitemOpen
  \bibfield  {author} {\bibinfo {author} {\bibfnamefont {T.~E.}\ \bibnamefont
  {Ouldridge}}, \bibinfo {author} {\bibfnamefont {A.~A.}\ \bibnamefont
  {Louis}}, \ and\ \bibinfo {author} {\bibfnamefont {J.~P.~K.}\ \bibnamefont
  {Doye}},\ }\href@noop {} {\bibfield  {journal} {\bibinfo  {journal} {J.
  Phys.: Condens. Matter}\ }\textbf {\bibinfo {volume} {22}},\ \bibinfo {pages}
  {104102} (\bibinfo {year} {2010})}\BibitemShut {NoStop}%
\bibitem [{\citenamefont
  {Ouldridge}(2012{\natexlab{b}})}]{Ouldridge_bulk_2012}%
  \BibitemOpen
  \bibfield  {author} {\bibinfo {author} {\bibfnamefont {T.~E.}\ \bibnamefont
  {Ouldridge}},\ }\href@noop {} {\bibfield  {journal} {\bibinfo  {journal} {J.
  Chem. Phys.}\ }\textbf {\bibinfo {volume} {137}},\ \bibinfo {pages} {144105}
  (\bibinfo {year} {2012}{\natexlab{b}})}\BibitemShut {NoStop}%
\bibitem [{\citenamefont {Dimitrov}\ and\ \citenamefont
  {Zuker}(2004)}]{Dimitrov2004}%
  \BibitemOpen
  \bibfield  {author} {\bibinfo {author} {\bibfnamefont {R.~A.}\ \bibnamefont
  {Dimitrov}}\ and\ \bibinfo {author} {\bibfnamefont {M.}~\bibnamefont
  {Zuker}},\ }\href@noop {} {\bibfield  {journal} {\bibinfo  {journal}
  {Biophys. J.}\ }\textbf {\bibinfo {volume} {87}},\ \bibinfo {pages} {215}
  (\bibinfo {year} {2004})}\BibitemShut {NoStop}%
\bibitem [{\citenamefont {Zhang}\ and\ \citenamefont
  {Winfree}(2009)}]{zhang2009control}%
  \BibitemOpen
  \bibfield  {author} {\bibinfo {author} {\bibfnamefont {D.~Y.}\ \bibnamefont
  {Zhang}}\ and\ \bibinfo {author} {\bibfnamefont {E.}~\bibnamefont
  {Winfree}},\ }\href@noop {} {\bibfield  {journal} {\bibinfo  {journal} {J.
  Am. Chem. Soc.}\ }\textbf {\bibinfo {volume} {131}},\ \bibinfo {pages}
  {17303} (\bibinfo {year} {2009})}\BibitemShut {NoStop}%
\end{thebibliography}%
\bibliographystyle{apsrev4-1}

\appendix

\section{oxDNA Model Details}
\label{app:model}

\begin{figure}
\centering
\includegraphics[width=0.45\textwidth]{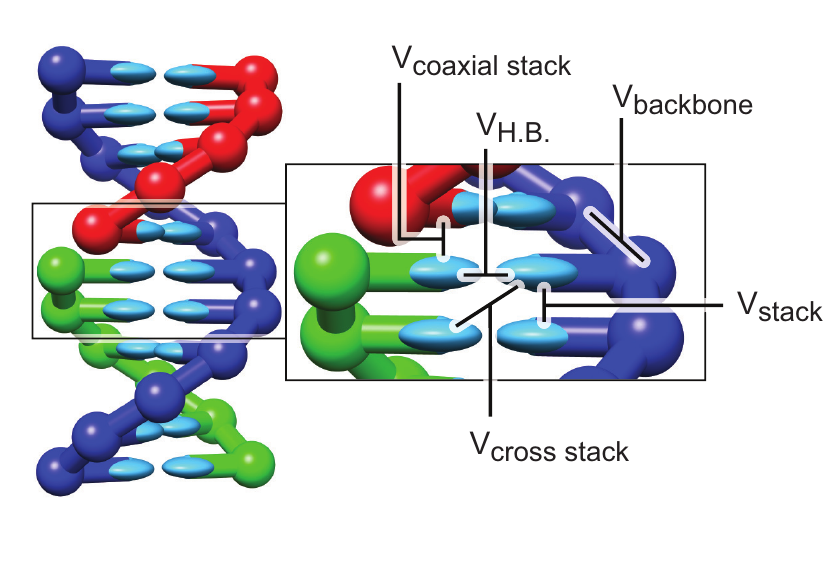}
\centering \caption{\footnotesize A model DNA duplex, with stabilizing interactions depicted schematically. The backbone sites are shown as spheres, the bases as ellipsoids. Backbone coloring indicates strand identity. All nucleotides also interact with repulsive excluded volume interactions. The coaxial stacking interaction acts like a stacking interaction between bases that are not immediate neighbors along the backbone of a strand. Taken from Ref.\,\onlinecite{Ouldridge_walker_2013}.
\vspace{-0.1in}}
\label{fig_interactions}
\end{figure}

OxDNA and its interaction potentials have been described
in detail elsewhere.\cite{Ouldridge2011,Ouldridge_thesis,Sulc2012}
The model represents DNA as a string of nucleotides, where each nucleotide
(sugar, phosphate and base group) is a rigid body with three interaction sites. The potential energy of
the system can be decomposed as 
\begin{eqnarray}
\label{eq_hamiltonian}
  V =  \sum_{\left\langle ij \right\rangle} \left( V_{\rm{b.b.}} + V_{\rm{stack}} +
V^{'}_{\rm{exc}} \right) + \nonumber \\
    \sum_{i,j \notin {\left\langle ij \right\rangle}} \left( V_{\rm HB} +  V_{\rm{cr.st.}}  +
V_{\rm{exc}}  + V_{\rm{cx.st.}} \right) ,
\end{eqnarray}
where the first sum is taken over all nucleotides that are nearest
neighbors on the same strand and the second sum comprises all remaining
pairs. The interactions between nucleotides are schematically shown in 
Fig. \ref{fig_interactions}.  The backbone potential $ V_{\rm{b.b.}}$ is an isotropic
spring that imposes a finite maximum distance between backbone sites of neighbors, mimicking
the covalent bonds along the strand. The hydrogen bonding ($V_{\rm HB}$), cross
stacking ($V_{\rm{cr.st.}}$), coaxial stacking ($V_{\rm{cx.st.}}$) and
stacking interactions ($V_{\rm{stack}}$) are anisotropic and explicitly depend on
the relative orientations of the nucleotides as well as the distance
between the relevant interaction sites. This orientational dependence captures the planarity of bases, and helps drive the formation of helical duplexes. The coaxial stacking term is
designed to capture stacking interactions between bases that are not immediate neighbors along the backbone of a strand. Base and backbone sites also have excluded volume interactions
$V_{\rm{exc}}$ and $V^{'}_{\rm{exc}}$.

Hydrogen-bonding interactions are only possible between complementary (A-T and C-G) base pairs. In the sequence-dependent parameterization that we use for all simulations, the strengths of interactions $V_{\rm{stack}}$ and $V_{\rm{HB}}$ further depend on the identity of the bases involved.\cite{Sulc2012}  Interactions were fitted to reproduce melting temperatures and transition widths of oligonucleotides, as predicted by SantaLucia's nearest-neighbor model.\cite{SantaLucia1998}
Note that our approach is significantly more complex than the nearest-neighbor model. We simply treat the latter as a high-quality fit to experimental data. For the purpose of parametrization, structural and mechanical properties of both double- and single-stranded DNA are also taken into account in the fitting procedure. In DNA the double helical structure emerges because there is a length-scale mismatch between the preferred inter-base distance along the backbone, and the optimal separation of bases when stacking. It is exactly this feature that drives the helicity of oxDNA, rather than an imposed natural twist on the backbone. Overall, the emphasis in our derivation of oxDNA was on physics relevant to the duplex formation transition. As discussed in the main text, oxDNA has been extensively tested for other DNA properties and systems to which it was not fitted. Our success in describing all these phenomena gives us confidence to use it to study the dynamics of hybridization in the presence of hairpins.

OxDNA was fitted to reproduce DNA behavior at $[\mbox{Na}^{+}] =
0.5$M, where the electrostatic properties are strongly screened, and it
is reasonable to incorporate them into a short-ranged excluded volume. The model therefore contains no further explicit electrostatic interactions. It should be noted that oxDNA neglects several features of DNA structure and interactions due
to the high level of coarse-graining. Specifically, the double helix in the
model is symmetrical; the grooves between the backbone sites do not have different sizes (i.e., major and minor grooving), and all four nucleotides have the
same structure. These differences with real DNA mean that oxDNA will not be able to treat phenomena that depend sensitively, for example, on anisotropic elasticity, explicit salt ion effects, or the existence of major and minor grooving. However, these specific properties of DNA are unlikely to be critical to the general arguments we are making about hybridization in the presence of hairpins in this article. Rather, it is the correct treatment of the basic mechanical properties of both single and double strands, together with the basic physics of hydrogen bonding and stacking that determines the emergent physical phenomena we are trying to describe.

\section{Details of hairpin systems}

\begin{table}
\begin{center}
\begin{tabular}{ l | l }
${\rm P}_0$ & 3$^{\prime}-$\text{GAG ACT TGC CAT CGT AGA ACT GTT G}$-$5$^{\prime}$\\
${\rm P}_3$ & 3$^{\prime}-$\text{TGA CGA TCA T\underline{GT C}TG CGT \underline{GAC} TAG A}$-$5$^{\prime}$\\
${\rm P}_4$ & 3$^{\prime}-$\text{A\underline{CA CG}A TCA TGT CTG \underline{CGT G}AC TAG A}$-$5$^{\prime}$\\
${\rm T}_0$ & 3$^{\prime}-$\text{CAA CAG TTC TAC GAT GGC AAG TCT C}$-$5$^{\prime}$\\
${\rm T}_3$ & 3$^{\prime}-$\text{TCT A\underline{GT C}AC GCA \underline{GAC} ATG ATC GTC A}$-$5$^{\prime}$\\
${\rm T}_4$ & 3$^{\prime}-$\text{TCT AGT \underline{CAC G}CA GAC ATG AT\underline{C GTC} A}$-$5$^{\prime}$\\ 
\end{tabular}
\caption{Sequences from Ref.~\onlinecite{Gao2006} used in this work. Underlined sequences show hairpin stems that are intended to form by design.}
\label{sequences}
\end{center}
\end{table} 

To compare simulations with experimental data, we simulated complementary strands, (Table~\ref{sequences}) for which association rate constants have been experimentally determined by Gao {\it et al.}~\cite{Gao2006}.  Gao {\it et al.} used UV absorbance spectroscopy to measure association rate constants for 3 pairs of complementary probe/target (P/T) strands that they had designed to exhibit varying degrees of secondary structure: no hairpins (P$_0$,T$_0$), 3-base-pair stem hairpins (P$_3$,T$_3$), and 4-base-pair stem hairpins (P$_4$,T$_4$). Strands with the same subscript are complementary, so if mixed together will form a duplex: P$_n$ + T$_n$ $\rightarrow$ P$_n$ T$_n$.
Their experiments were performed at room temperature (\ang{20}C), a ssDNA concentration of \SI{2}{\micro\Molar}, and high salt concentration of [Na$^+$] = \SI{0.5}{\Molar}. This high salt concentration~\cite{Yakovchuk2006} is the same as the duplex melting temperatures to which oxDNA has been parameterized,\cite{Ouldridge2011} so duplex behavior should be reproduced well. At this salt concentration, electrostatic interactions between nucleotides are screened. These electrostatic interactions oppose DNA structure formation, so DNA nanotechnology usually uses similar high salt concentrations.

\section{Simulation Methods}
\label{sim_methods}

\subsection{Thermodynamics}
 
\subsubsection{Virtual Move Monte Carlo}

A standard approach for calculating thermodynamic properties of computational models is the Metropolis algorithm.\cite{Frenkel2001} A drawback with this approach is that only moving single particles at a time results in slow equilibration for systems with strong attractions. This is true for DNA strands, where collective diffusion is strongly suppressed if nucleotides are moved individually. Simulations can be made more efficient by using the Virtual Move Monte Carlo (VMMC) algorithm proposed by Whitelam and Geissler which allows for collective diffusion using cluster moves of particles.\cite{whitelam2009role}. Specifically, we use the variant presented in the appendix of Ref.~\onlinecite{whitelam2009role}. Initially a particle is selected, and a move is chosen at random as in the Metropolis algorithm. The particle's neighbors are then added to a co-moving`cluster' with probabilities determined by the energy changes that would result from the move. Consequently, multiple particles tend to move at once. To use VMMC, we must select `seed' moves of a single particle. For all VMMC simulations reported here, the seed moves were:
\begin{itemize}
\item Rotation of a nucleotide about its backbone site, with the axis chosen uniformly on the unit sphere and the angle drawn from a normal distribution with a mean of zero and a standard deviation of 0.22 radians.\\
\item Translation of a nucleotide, where the displacement along each Cartesian axis is drawn from a normal distribution with a mean zero and a standard deviation of 0.15 simulation units of length (0.1277 nm).
\end{itemize}
To improve efficiency, if the algorithm generates a cluster move involving more than 7 particles the move is automatically rejected.

\subsubsection{Umbrella Sampling}

An important concept is that of a reaction coordinate (or order parameter) $Q$, which groups together microstates of a system that share some macroscopic property (for example, all configurations of strands with a certain number of base pairs). The free-energy profile as a function of $Q$ can provide useful information about the reaction, provided an appropriate choice has been made. In particular, free-energy barriers can make certain regions of configuration space hard to reach, which prevents efficient sampling of all of the states of interest. The free-energy landscape can be artificially flattened by weighting states with different values of $Q$ appropriately, a technique known as umbrella sampling.\cite{torrie1977nonphysical} Thermodynamic properties of the system can then be extracted from simulations by unweighting the resulting distributions.

In particular, for an unweighted simulation a particular microstate with coordinates $\textbf{q}^N$ and energy $E(\textbf{q}^N)$ is sampled with probability 
\begin{equation}
P(\textbf{q}^N) \propto e^{-\beta E(\textbf{q}^N)}.
\end{equation}
The equilibrium average of some variable $A(\textbf{q}^N)$ is then given by the sum over all states, weighted by their Boltzmann factors:
\begin{equation}
\langle A \rangle = \frac{ \int A(\textbf{q}^N) e^{-\beta E(\textbf{q}^N)} d\textbf{q}^N }{\int e^{-\beta E(\textbf{q}^N)} d\textbf{q}^N }.
\label{average_quantity}
\end{equation} 
By applying a weighting $w = w(Q(\textbf{q}^N))$ to each value of the order parameter, we change the sampling frequency to 
\begin{equation}
P_{w}(\textbf{q}^N) \propto w(Q(\textbf{q}^N)) e^{-\beta E(\textbf{q}^N)}.
\end{equation}
where the subscript $w$ indicates a property of the weighted system. So we can artificially ensure that our simulation samples all states equally by making $P_w$ constant for all microstates. Equilibrium thermodynamic properties are then obtained by unbiasing afterwards, as can be seen by rewriting Eq.~(\ref{average_quantity}) as follows:
\begin{eqnarray}
\langle A \rangle &=& \frac{ \int \frac{A(\textbf{q}^N)}{w(Q(\textbf{q}^N))} w(Q(\textbf{q}^N)) e^{-\beta E(\textbf{q}^N)} d\textbf{q}^N }{\int \frac{1}{w(Q(\textbf{q}^N))} w(Q(\textbf{q}^N)) e^{-\beta E(\textbf{q}^N)} d\textbf{q}^N } \\ 
&=& \frac{ \langle A / w \rangle_w }{ \langle 1/w \rangle_w}.
\label{weighted_average_quantity}
\end{eqnarray} 
Throughout this article, it makes sense to use the number of base pairs in our definition of $Q$. This is the usual choice for studying hybridization processes.

\subsubsection{Single Histogram Reweighting}


To determine melting temperatures for structures, we implemented the temperature extrapolation method known as single histogram reweighting, based on the method introduced by Ferrenberg and Swensden.\cite{Ferrenberg1988} The method of single histogram reweighting allows extrapolation of results from simulations at a particular temperature $T_0$ to other temperatures.\cite{Ferrenberg1988}  States of the system are grouped by their value of some quantity $A$ and their energy $E$, so a histogram $p(A, T_0, E)$ can be produced. The temperature-independent density of states $\Omega(A,E)$ can then be inferred via
\begin{equation}
p(A,T_0,E) \propto \Omega(A,E) e^{-\beta_0 E}
\label{extrap_prob}
\end{equation}
where $\beta_0 = 1/R T_0$. The proportionality constant is unknown because we can only ever know the relative ratios of states in our simulations. $\Omega(A, E)$ can then be used to calculate the average value of $A$ at any temperature $T$ by integrating over all possible states:
\begin{equation}
\label{histo1}
\langle A(T) \rangle = \frac{\int \int A \Omega(A,E) e^{-\beta E} dEdA}{\int \int \Omega(A,E) e^{-\beta E} dEdA}.
\end{equation}
We can rewrite this using Eq.~(\ref{extrap_prob}) as
\begin{equation}
\label{histo2}
\langle A(T) \rangle = \frac{\int \int A p(A,T_0,E) e^{-(\beta_0 - \beta) E} dEdA}{\int \int p(A,T_0,E) e^{-(\beta_0 - \beta) E} dEdA}.
\end{equation}
We point out that our potential energy function (Eq.~(\ref{eq_hamiltonian})) depends explicitly on the temperature through $V_{stack}$. It is straightforward to extend Eqs.~(\ref{histo1}) and (\ref{histo2}) to our case. It must be pointed out that the extrapolation can only go so far from the temperature $T_0$ because sampled states will not be representative of the dominant states at other temperatures. For this reason we explicitly calculate free energies at $T = $ \ang{20}C for the main data of the manuscript and in subsequent sections here. We only use histogram reweighting for melting curves in the vicinity of the $T_m$ for a particular system, which are of less importance. 

In all VMMC simulations employing umbrella sampling (for single strand and double strand systems), we estimated the error of the computed relative free energies corresponding to different states by computing the standard error of the mean value of multiple independent simulations. The details of each simulation, including the order parameters used as well as the number of independent simulations, are discussed in Section~\ref{sim:thermo}. 

\subsection{Kinetics}
\subsubsection{Molecular Dynamics}
\label{sec:dynamics}
Kinetic simulations were performed using an Anderson-like thermostat, similar to the one described in appendix A of Ref.~\onlinecite{russo2009reversible}. The Newtonian equations of motion for the system are integrated by Verlet integration~\cite{verlet1967computer} with a discrete time-step $\delta t$, so that the positions, velocities, orientations, and angular velocities of the nucleotides are recalculated at each time-step. This alone would give the DNA strands constant energy and cause ballistic motion. In reality, DNA in a solvent is being bombarded by water particles and thus undergoes Brownian motion. To model Brownian motion, the velocity of each nucleotide is resampled with a probability $p_{v} = 0.02$ from a Maxwell-Boltzmann distribution at the temperature of the solvent every 103 time steps. The algorithm also resamples angular velocities with a different probability $p_{\omega} = 0.0067$. The solvent thus acts as a large heat bath at a fixed temperature, ensuring that the simulated system samples from the canonical ensemble. On time scales longer than $N_{Newt}\delta t/p_v$, where $\delta t$ is the integration time step, the dynamics is diffusive. Choosing $\delta t$ = \num{3.03e-12}s for all dynamics simulations in this study gives a diffusion constant $D_{sim}$ that is about 100 times higher than experimental measurements~\cite{Lapham1997} of $D_{exp}$ = \num{1.19e-10}\SI{}{\metre\squared\per\second}.

This is a common procedure for coarse-grained models where higher diffusion constants can be used to accelerate diffusion. Accelerated diffusion can also speed up certain processes by smoothing out, on a microscopic scale, energy profiles.\cite{Murtola2009} This can be advantageous because it means simulations utilizing coarse-grained models can be used to study more complicated systems. In a previous study using oxDNA, the hybridization kinetics of a non-repetitive sequence was considered.\cite{Ouldridge_binding_2013} In that study it was shown that using higher friction constants (smaller diffusion coefficients) in simulations utilizing Langevin dynamics at 300K slowed down hybridization, but did not otherwise qualitatively affect the results. In particular, the tendency for initial base pairs to melt away rather than lead to a full duplex was found to be preserved. Our systems are similar to those studied in Ref.~\onlinecite{Ouldridge_binding_2013}, possessing similar numbers of total base pairing between the strands, and using a similar simulation temperature. Additionally, many approximations of real DNA have already been made in the construction of the oxDNA model, and we expect that running simulations with a diffusion coefficient that is larger than the experimentally measured value should preserve the effects that hairpins in single strands have on the relative hybridization and dissociation rates. 


\subsubsection{Forward Flux Sampling}

\begin{figure}
\begin{center}
\includegraphics[width=225pt]{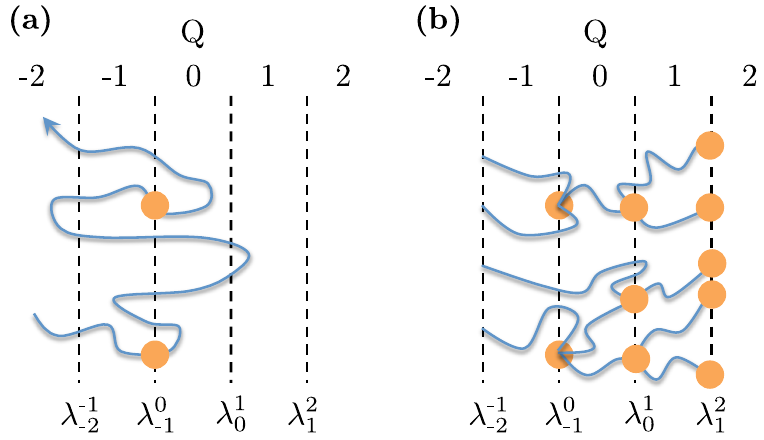}
\caption{\footnotesize (Color online) (a) Schematic illustration of the interfaces involved in flux generation. The flux is initially measured across the interface $\lambda_{-1}^{0}$. The orange dots indicate that a crossing by a trajectory contributes to the flux. These are also the states that are used to launch successive stages of the simulation. (b) In direct FFS, large numbers of configurations are randomly selected from the set that successfully crossed the interface $\lambda_{-1}^{0}$, and the probability of crossing the interface $\lambda_{0}^{1}$ is measured before any configuration goes back to the interface $\lambda_{-2}^{-1}$. The process is then iterated over successively chosen interfaces until reaching $Q_{\rm max}$. Figure adapted from Ref.~\onlinecite{Ouldridge_binding_2013}. }
\label{FFS}
\end{center}
\end{figure}

`Brute force' dynamics simulations using an Anderson-like thermostat are not efficient enough to generate a representative ensemble of transitions that start from diffusing single strands and hybridize into the 25 base pair duplex in a reasonable amount of time. Thus, we resorted to using Forward Flux Sampling (FFS) to more efficiently calculate fluxes between local free-energy minima as well as sample the transition pathways. The term `flux' from (meta)stable state A to state B has the following definition.

\vspace{0.2cm}
\begin{adjustwidth}{1.0cm}{1.0cm}
Given an infinitely long simulation in which many transitions are observed, the flux of trajectories from A to B is $\Phi_{AB} = N_{AB} / (\tau f_A)$, where $N_{AB}$ is the number of times the simulation leaves A and then reaches B, $\tau$ is the total time simulated, and $f_A$ is the fraction of the total time simulated for which state A has been more recently visited than state B.
\end{adjustwidth}
\vspace{0.2cm}

FFS requires use of an order parameter, $Q$, which provides a descriptive measure of the extent of the reaction between states A and B. Additionally, the order parameter must be chosen such that non-intersecting interfaces $\lambda_{Q-1}^{Q}$ can be drawn between consecutive values of $Q$. At the beginning of an FFS implementation, a brute force simulation is run starting from states described by $Q$ = -2, and the flux of trajectories crossing the surface $\lambda_{-1}^{0}$ is measured. The total flux of trajectory from $Q$ = -2 to another free-energy minimum $Q = Q_{max}$ can be calculated as the flux of trajectories crossing $\lambda_{-1}^{0}$ multiplied by the probability that trajectories reach $Q = Q_{\rm max}$, all before returning to $Q$ = -2. When the order parameter is partitioned into $N$ number of surfaces to be crossed before reaching $Q_{\rm max}$, but before returning to $Q$ = -2, it can be factorized as 
\begin{equation}
P\left( \lambda_{Q_{\rm max}-1}^{Q_{\rm max}} | \lambda_{-1}^{0} \right) = \prod_{Q=1}^{Q_{\rm max}} P\left( \lambda_{Q-1}^{Q} | \lambda_{Q-2}^{Q-1} \right).
\end{equation}
Next, the simulation loads random configurations that have just crossed $\lambda_{-1}^{0}$, which are used to estimate $P(\lambda_{0}^{1}|\lambda_{-1}^{0})$. The process is then iterated for successive interfaces, and the flux as well as the trajectories that successfully reach $Q_{max}$ from the distribution of pathways can be measured. 

We estimated the random error in the FFS simulations in the following way. In Table~1 in the main text we report the mean value for the hybridization rates from 5 identical and independent FFS simulations for each system. The error reported for each system is the standard error of the mean value. In Tables~\ref{p0t0_ffs},~\ref{p3t3_ffs}, and~\ref{p4t4_ffs}, we report the mean and the standard error of the mean of for each individual interface. Each calculation of the flux in the three systems was repeated 240 times in total (48 simulations were used to obtain the flux in each independent calculation of the rate), while the probability of crossing interface $\lambda_{Q-1}^Q$ was computed from the 5 independent calculations of the rate. 

\section{Simulation Protocols}
\label{sim:general}
In this section we discuss the implementation of the algorithms of Section~\ref{sim_methods} for both single-stranded and duplex systems. As mentioned in Section~\ref{sec:dynamics}, we simulated the three duplex systems using molecular dynamics and VMMC simulations. Unless otherwise stated, the temperature in a simulation was taken to be $T$ = \ang{20}C, which is the same temperature used by Gao {\it et al.} in the experiments. Additionally, for simulations of the duplex systems, we used a simulation box with a volume of \num{3.96e-23}\SI{}{\cubic\metre} which corresponds to a concentration of \SI{42}{\micro\Molar}, and is 21 times larger than the experimental concentrations of \SI{2}{\micro\Molar} used for each system. We also used two types of order parameters in the simulations that can be used to construct multi-dimensional order parameters, which are discussed specifically for each simulation in the sections to follow. In particular, a `distance order parameter' measures the minimum distance between hydrogen-bonding sites over correct pairs of bases in the two strands. A `bonds' order parameter measures the total number of base pairs, which can be specified to be intra- or inter-strand base pairs. The definition of a bonded base pair in our simulations is two bases with a hydrogen bonding energy below \SI{0.596}{\kilo\cal\per\mole}.  This value for the selected cutoff corresponds to about 15\% of typical hydrogen-bond energy. 

\subsection{Thermodynamics}
\label{sim:thermo}

\subsubsection{Single-strand Thermodynamics}

Melting properties for ``monomers" (secondary structure of isolated strands) can be calculated from $\Phi$, the ratio of bound to unbound states in a simulation of a single strand. For self-interacting strands, the fraction of folded states in a hypothetical bulk system is concentration-independent and can be inferred from
\begin{equation}
f^{\rm mon}_{\rm bulk} = f_{\rm sim} = \frac{\Phi}{1+\Phi}.
\label{mon_conc}
\end{equation}
The melting temperature is taken to be the point where $f^{\rm mon}_{\rm bulk} =1/2$. 

We determined the approximate location of the melting temperatures of the hairpins to ensure that they were stable at $T$ = \ang{20}C. oxDNA has been shown to reproduce the dependence of hairpin melting temperature on stem-length and loop-length.\cite{Ouldridge2011}. We ran 10 independent VMMC simulations with \num{6.3e+10} steps at temperatures of $T$ = \ang{20}C for both the P$_0$ and T$_0$ strands, \num{1.1e+11} and \num{8.8e+10} steps at $T$ = \ang{45}C for the P$_3$ and T$_3$ strands respectively, and \num{1.1e+11} steps at $T$ = \ang{60}C for the P$_4$ and T$_4$ strands. Defining melting temperatures of hairpins is complicated because strands may exist in multiple stable structures. We were interested in the point where the strands did not have significantly stable intra-strand base pairs, so we counted all states with at least one intra-strand base pair as `bound' and states with no intra-strand base pairs as `unbound', then calculated the yields from Eq.~\ref{mon_conc}. The yield curves for P and T strands are obtained by single histogram reweighting and are shown in Fig.~\ref{hairpin_yield}, and in Fig.~\ref{yield} in the main text for just the P strands. 

In addition to the melting curves, we also determined free-energy landscapes at $T$ = \ang{20}C from VMMC simulations with umbrella sampling for the P$_0$, P$_3$, and P$_4$. For P$_0$ strands we used an order parameter that kept track of any intra-strand base pair. For the P$_3$ and P$_4$ strands, we used a two-dimensional order parameter where the two coordinates describe (1) the intended base pairs according to Gao {\it et al.}\cite{Gao2006} (`correct' base pairs) of the structures, and (2) all possible other intra-strand base pairs (total base pairs minus correct base pairs). For a typical strand there was $\sim$80 possible different intra-strand base pairs. The free energies for the strands were calculated from cumulative distributions from 10 parallel runs with \num{5e+10}, \num{5.5e+10}, and \num{2.5e+10} steps for P$_0$, P$_3$ and P$_4$, respectively. The results for the free energies and yields of the single strands are shown in Figs.~\ref{hairpin_free} and \ref{hairpin_yield}, respectively, and are discussed in Section~\ref{strand_free}.

\subsubsection{Duplex Thermodynamics}

We first computed the melting temperatures of the three duplex systems. For structures consisting of two molecules care must be taken in extrapolating from a simulation of two strands to a bulk solution with many more strands, because fluctuations in local concentrations play an important role. If $\Phi$ is the ratio of bound to unbound states in a simulation of two molecules, the yield of a non-self-complementary duplex in a bulk solution (with the same average concentration of reactants) is given in Ref.~\onlinecite{Ouldridge_bulk_2010} as
\begin{equation}
f^{\rm dim}_{\rm bulk} = \left(1+\frac{1}{2 \Phi}\right) - \sqrt{\left(1+\frac{1}{2 \Phi}\right)^{2} - 1}.
\label{dimer_conc}
\end{equation}
The melting temperature occurs when $f^{\rm dim}_{\rm bulk}$, which corresponds to a simulation yield of $\phi=2$. To compare simulations of single duplexes with experimental data, as in Table~\ref{melting_temps} for the melting temperatures of the 3 systems, $\Phi$ was measured in simulations and then scaled to the experimental concentration ($\Phi$ is proportional to the concentration, so scales from a concentration c$_1$ to another concentration c$_2$ by the factor c$_2$/c$_1$). The bulk yield was then calculated by using Eq.~(\ref{dimer_conc}). Note that this approximation only works if the systems are essentially ideal -- the accuracy of this approximation has been previously established for oxDNA under similar conditions~\cite{Ouldridge_bulk_2010,Ouldridge_bulk_2012}. We ran 10 VMMC simulations with umbrella sampling for each system using an order parameter that measured the number of inter-strand bonds between the two strands. The simulations for the three systems were carried out at $T =$ \ang{77.5}C, which is near the melting temperature of each system. For P$_0$T$_0$, P$_3$T$_3$, and P$_4$T$_4$, each of 10 simulations ran for \num{1.6e+10} steps, \num{2.9e+10} steps, and \num{1.8e+10} steps, respectively. The results for the duplex yields are shown in Fig.~\ref{duplex_yields} and discussed in Section~\ref{duplex_free_energies}.

Next, the computations of the relative free energies of the P$_0$T$_0$, P$_3$T$_3$, and P$_4$T$_4$ systems at $T =$ \ang{20}C were carried out using VMMC moves along with umbrella sampling with a multi-dimensional order parameter that measures (1) the number of intra-strand base pairs in the P strand, (2) the number of intra-strand base pairs in the T strand, and (3) the number of inter-strand base pairs between P and T strands. In the order parameter, any complementary bond is taken into account. This means we include all secondary structural base pairs in the single strands. We ran 10 simulations for \num{2.3e+11} steps for the P$_0$T$_0$ system and for \num{1.1e+11} steps for the P$_3$T$_3$ system. The results are shown in Figs.~\ref{p3t3_free}(a) and (b) for P$_0$T$_0$ and P$_3$T$_3$, respectively, and discussed in Section~\ref{duplex_free_energies}.

The same order parameter used for P$_0$T$_0$ and P$_3$T$_3$ was initially used for P$_4$T$_4$. Ten simulations utilizing umbrella sampling each ran for \num{1.1e+11} VMMC steps. These results are shown in Fig.~\ref{free_energy}(b) in the main text. During the course of the simulations, we noticed a difficulty in sampling the `pseudoknotted' intermediate states (in which the strands were bound by both tails and loops of the hairpins, illustrated in Fig.~\ref{pseudoknot}) that were observed in kinetic simulations; the order parameter was not efficient in driving their formation. We sampled these intermediate states in separate simulations with a distinct multi-dimensional order-parameter depending on (1) only the intended 4-stem hairpins in the P strand and (2) in the T strand, (3) the number of base pairs between the loops of the hairpins, (4) the number of base pairs between the two strands not including the loop-loop base pairs, and (5) the correctly aligned duplex base pairs between the strands. We sampled only those states in which at least one intended hairpin base pair was present in each strand, and also only states where coordinate (4) had at least one base pair formed. These simulations allow an estimate of the free energy of the pseudoknotted state compared to the tail-bound state. We ran a set of 10 simulations using this order parameter for \num{2e+11} VMMC steps each. These results are shown in Fig.~\ref{p4t4_free}(b) and are discussed in Section~\ref{duplex_free_energies}.

\subsection{Kinetics}

\subsubsection{FFS Simulation Details}

In this section we discuss the implementation of the FFS algorithm, discussed in Section~\ref{sim_methods}. As mentioned in Section~\ref{sec:dynamics}, we simulated the three duplex systems using molecular dynamics at the experimental temperature of $T$ = \ang{20}C. Additionally, in all kinetics simulations we used a simulation box with a volume of \num{3.96e-23}\SI{}{\cubic\metre} which corresponds to a concentration of \SI{42}{\micro\Molar} that is 21 times larger than the experimental concentration, as was noted in Section~\ref{sim:general}. We also use the same definition of a bonded base pair that was discussed in Section~\ref{sim:general}. 

\subsubsection{Order Parameter Used in FFS Simulations}

\begin{table}
    \renewcommand{\arraystretch}{1.5}
    \begin{tabular}{>{\centering}m{1cm} c }
    \hline
    $Q$ & Description \\
    \hline
    -2 & $d >$ \SI{5.1}{\nano\metre} \\ 
    -1 & \SI{1.7}{\nano\metre} $<  d \le$ \SI{5.1}{\nano\metre} \\
     0 & \SI{1.02}{\nano\metre} $< d \le$ \SI{1.7}{\nano\metre} \\
     1 & \SI{0.57}{\nano\metre} $< d \le$ \SI{1.02}{\nano\metre} \\
     2 & $d \le$ \SI{0.57}{\nano\metre} \& $x = 0$\\ 
     3 & $x \ge 1$ \\
     4 & $y \ge 2$ \\
     5 & $y \ge 6$ \\
     6 & $y \ge 15$ \\
     7 & $y = 25$ \\
    \hline\hline
    \end{tabular}
    \caption{The order parameter used in FFS simulations of duplex hybridization in all three systems studied. The parameter $d$ is the minimum distance between any intended base pair on the two strands, $x$ is the number of inter-strand base pairs between the two strands which includes mis-aligned and intended duplex base pairs, while $y$ is only the number of intended duplex base pairs between the two strands. For both $x$ and $y$, a base pair is taken to be present if the hydrogen-bonding energy is less than \SI{-0.596}{\kilo\cal\per\mol}. }
    \label{op_ffs}
\end{table}

The order parameter used in simulations is detailed in Table~\ref{op_ffs}. Specifically, we use a combination of distance and bond criteria as outlined in Section~\ref{sim:general}. Distance criteria are used to define states $Q=-2 \rightarrow 2$, and bonding criteria for states $Q=2 \rightarrow 7$. 
For the $Q=2,3$ states, we allowed the bond criteria to track any inter-strand bond between the two strands, which allowed us to monitor the number of non-intended inter-strand base pairs (i.e. mis-aligned base pairs) and also the number of correctly aligned inter-strand base pairs that have formed during the initial association events. The bond criteria for states $Q=4-7$ track only correctly aligned inter-strand base pairs.

\subsubsection{Initial Equilibration of Single-strand States}


Before implementing FFS, we performed lengthy equilibration simulations to ensure that the single strands were initialized in thermodynamically representative states. Here we describe the procedure used to select these states. In Section~\ref{strand_free} the relative free energies for each single strand were computed using VMMC with umbrella sampling. We performed similar simulations except that both P and T strands were simulated in the same box corresponding to a concentration of \SI{42}{\micro\Molar}. A 3-dimensional order parameter was used that measures (1) the minimum distance between any pairs of nucleotides that are intended to be base pairs in the final duplex, (2) the number of intra-strand base pairs in the P strand, and (3) the number of intra-strand base pairs in the T strand. The strands were prevented from coming within \SI{5.1}{\nano\metre} of each other, as measured by coordinate (1). We ran 10 simulations for \num{5e9} VMMC steps and saved configurations every \num{5e6} VMMC steps, which ensured that any two saved configurations were energetically decorrelated from each other. In total we collected a set containing 10000 configurations. For each state described by the 3-dimensional order parameter there is an umbrella bias $w(Q)$. We randomly selected a configuration from the set and saved it to be used in FFS simulations if $w(min)/w(Q) \le R$, where $R$ is a random number selected within the range $0<R\le 1$ and $w(min)$ was the smallest biasing weight applied in the simulation. This step was repeated until 200 configurations were obtained for each of 5 independent FFS simulations. At the start of each flux generation simulation an initial configuration from the saved set of 200 was selected at random and set to be the starting configuration.

\subsubsection{Second-order Kinetics Approximation}

An important question is whether or not the approximation of instantaneous reactions ({\it i.e., second-order kinetics}) is valid for oxDNA. Such an approximation is reasonable if the time taken from first interaction to full duplex formation or separation of strands is small compared to the diffusional time scale governing the first contact between strands. We did observe simulations spending significant computational times in states where the partially hybridized strands had formed kissing-hairpins. Theoretically, FFS should account for intermediates states with long lifetimes  during flux generation. However, as the formation of such states is rare during flux generation, the sampling is poor. As an alternative to the brute-force approach, we assumed second-order kinetics, reducing the sampling challenge during flux generation, and then checked the accuracy of the assumption from the resultant data.

During flux generation, we therefore restarted (from $Q=-2$) trajectories that reached $Q=3$, the first state in which a bond is present between strands. Consequently, any time spent in configurations with bonds between strands was not measured in our simulations. Technically our overall FFS protocol measured the flux from $Q=-2$ to $Q=2$, and the subsequent probability of reaching $Q=Q_{\rm max}$ before returning to $Q=-2$. This approximates the flux from $Q=-2$ to $Q=Q_{\rm max}$ provided that the time spent in the intermediate states is small -- in this limit, the measured flux is also proportional to the second-order rate constant.


To justify this assumption, we measured the time taken for shooting trajectories launched from intermediate values of $Q = Q^\prime$ to reach $Q = Q^\prime +1$ or $Q=-2$. We could therefore  determine the typical time taken for a configuration starting in the state $Q^\prime$ to either rearrange and proceed to a full duplex ($Q=Q_{\rm max}$), or to dissociate (taken to be when the system reaches the state $Q = -2$), for comparison with the diffusional time scale of first contact. These results are shown and discussed in Section~\ref{subtleties}. We find that the second-order approximation is reasonable at the concentration used in the simulations (and would be even better at the experimental concentration, 21 times lower) and therefore the relative fluxes estimated by our approach are decent predictions for the relative rate constants in oxDNA.

\section{Results}

\subsection{Thermodynamics}

\subsubsection{Single-strand Thermodynamics}
\label{strand_free}

Gao {\it et al.} used the mFold software~\cite{Dimitrov2004} to design the strands listed in Table 1, and assumed the predicted lowest energy structures to be the only important structures in their investigation.  mFold uses the nearest-neighbor thermodynamic model developed by Santa Lucia\cite{SantaLucia1998} to analyze secondary structure. However, this model cannot yet incorporate more complex structures like pseudoknots or multiple internal loops. They also cannot take into account forces that may result from the three-dimensional structure, which can be important in some cases~\cite{zhang2009control}. 

mFold predicts that at $T$ = \ang{20}C the P$_0$ T$_0$ strands, while designed to minimize their secondary structure, had multiple possible structures with free energies close to zero relative to the hairpin-free case, showing the difficulty of eliminating hairpins completely from long strands. From the simulations, we also found that the P$_0$ and T$_0$ strands were not dominated by any particular hairpin, but did frequently have some limited secondary structure. These transient base pairs should have a limited effect on hybridizationas they can melt easily.

\begin{figure}
\begin{center}
\includegraphics[width=225pt]{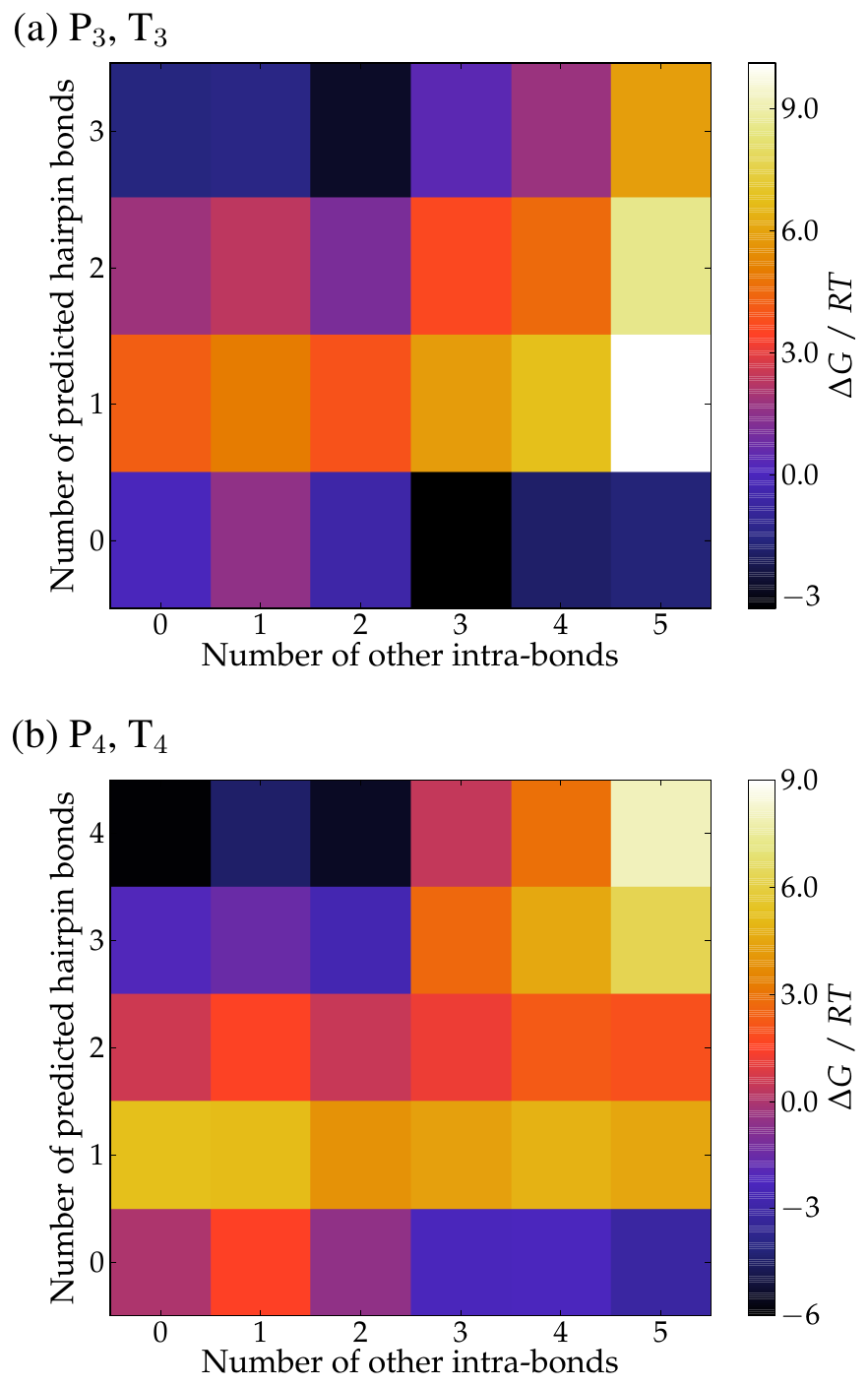}
\caption{(Color online) The free-energy profiles of the P$_3$ and P$_4$ strands are shown in (a) and (b), respectively. The free energy of a particular state relative to the unbound (0,0) state is indicated by the color of the square. The results for complementary T strands are essentially identical. }
\label{hairpin_free}
\end{center}
\end{figure}

\begin{figure}
\begin{center}
\includegraphics[width=225pt]{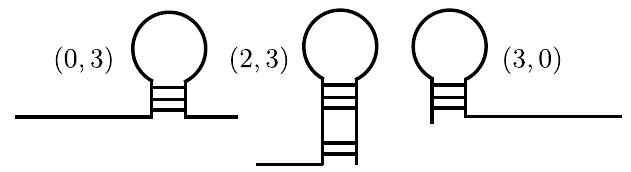}
\caption{Schematic representations of the significant hairpin states of the P$_3$ and T$_3$ strands. The $(0,3)$ state is the hairpin predicted by Nupack to be the most stable, the $(2,3)$ state is the predicted hairpin with the shorter tail folded in, and the $(3,0)$ state is a hairpin which appears to be slightly more stable than the predicted hairpin in oxDNA simulations, with the stem at the other end of the strand.}
\label{p3_hairpins}
\end{center}
\end{figure}

For the P$_3$ T$_3$ strands mFold predicted several structures with free energies within $\sim k_{B} T$ of the lowest free energy structure. This means they will be present in solution at significant concentrations. Free-energy profiles for these strands computed using oxDNA are plotted in Fig.~\ref{hairpin_free}. From Fig.~\ref{hairpin_free}(a) it is clear for the P$_3$ and T$_3$ strands that the predicted 3-base pair hairpin in the $(0,3)$ state is low in free energy compared to the state with no secondary structure. However, there exist other significant states of the strands, notably the $(3,0)$ state, which incorporates several 3-stem hairpins of a similar free energy. We found that this state was dominated by a hairpin with a very long 12-base pair tail. The $(4,0)$ and $(5,0)$ states showed pseudoknot behavior, with the strand bending back on itself twice. These states cannot be predicted by the model mFold uses. The $(2,3)$ state corresponded to the predicted 3-stem hairpin, but with the tails partially hybridized with two base pairs, producing a smaller trailing tail for the structure. These significant hairpin states are shown schematically in Fig.~\ref{p3_hairpins}. The P$_4$ and T$_4$ strands both show that the predicted 4-base pair hairpin in the $(0,4)$ state is extremely stable. The $(2,4)$ state consists of the predicted hairpin with two of the base pairs in the loops bonded.

\begin{figure}
\begin{center}
\includegraphics[width=225pt]{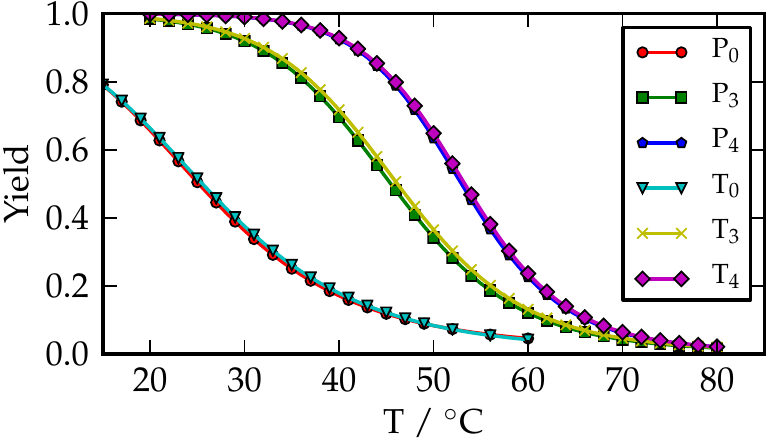}
\caption{Yields of all secondary structure of strands as a function of temperature. Error bars on all points are smaller than the symbol size. }
\label{hairpin_yield}
\end{center}
\end{figure}

To determine the prevalence of stable secondary structure at room temperature, we calculated the yields of the three P strands, which were plotted in Fig.~\ref{yield} in the main article. Comparable results for the T strands are shown in Fig.~\ref{hairpin_yield}, which show almost no difference between P and T strands. 

In summary, analysis of the secondary structure indicates that the hairpins are stable at room temperatures, and have significant and observable effects on the hybridization of strands. OxDNA and mFold disagree slightly when estimating the relative stabilities of similar structures. However, these subtleties are likely to be relatively unimportant, as both predict that hairpins are much more stable in P$_3$T$_3$ than P$_0$T$_0$, and also much more stable in P$_4$T$_4$ than in P$_3$T$_3$. Further, the main hairpin stem is always predicted to be of 3 and 4 base pairs in P$_3$T$_3$ and P$_4$T$_4$ respectively.

\subsubsection{Duplex Thermodynamics}
\label{duplex_free_energies}

\begin{figure}
\begin{center}
\includegraphics[width=225pt]{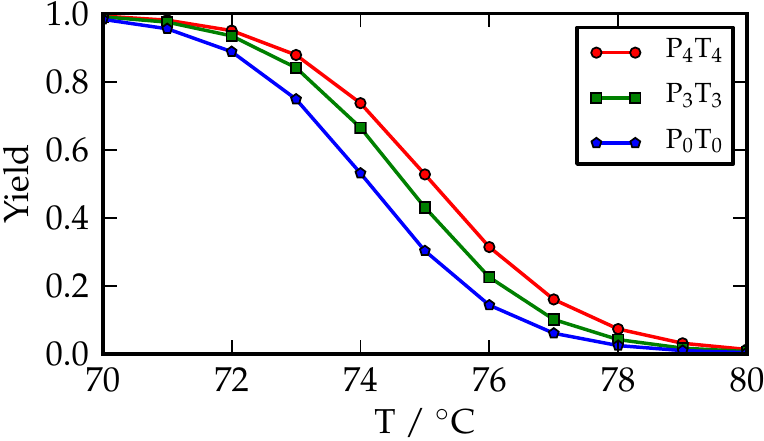}
\caption{Yields of all inter-strand structure of the three systems as a function of temperature. Error bars on all points are smaller than the symbol size used.}
\label{duplex_yields}
\end{center}
\end{figure}

\begin{table}
\begin{center}
\begin{tabular}{c|c|c|c}
Duplex & Gao {\it et. al}~\cite{Gao2006} & Santa Lucia~\cite{SantaLucia1998} & oxDNA \\
\hline
 P$_0$T$_0$ & 76.2 & 76.9 & 74.1 \\
 P$_3$T$_3$ & 77.4 & 77.4 & 74.6 \\
 P$_4$T$_4$ & 78.0 & 77.6 & 75.1 
\end{tabular}
\end{center}
\caption{Melting temperatures of P/T duplexes in Celsius at [ssDNA] = \SI{2}{\micro\Molar}, as measured experimentally in Ref.~\onlinecite{Gao2006}, calculated with the SantaLucia model~\cite{SantaLucia1998}, and simulated with oxDNA.}
\label{melting_temps}
\end{table}

Yield curves for bulk solutions in the region of the melting temperature are plotted in Fig.~\ref{duplex_yields} for the three duplex systems, and the melting temperatures are listed in Table~\ref{melting_temps} alongside the experimental values as measured by UV absorbance spectroscopy by Gao {\it et al.}~\cite{Gao2006}, and the predicted values that were calculated by the Santa Lucia model~\cite{SantaLucia1998}. All three methods are in agreement as to the order of the melting temperatures, although oxDNA appears to underestimate the true value by around \ang{3}C, corresponding to an error of  $<$1\%. This is unlikely to be of significance. What is important for this investigation is that all duplexes melt at temperatures well above room temperature, the order of stability is reproduced, and the differences between the curves is small. 

\begin{figure}
\begin{center}
\includegraphics[width=225pt]{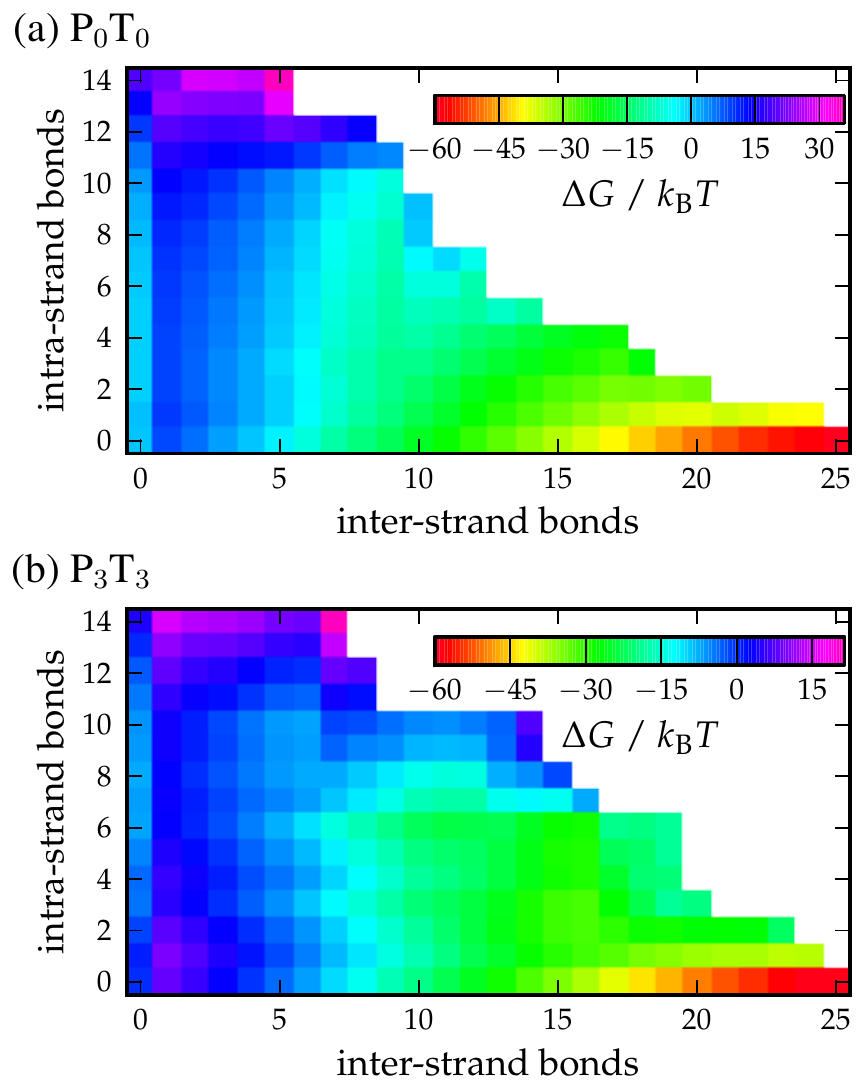}
\caption{(Color online) Free energies measured relative to the unbound state containing no intra-strand structure are plotted in (a) for P$_0$T$_0$ and in (b) for P$_3$T$_3$ as a function of any complementary inter-strand base pair present between the two strands, and any complementary intra-strand base pair present within P or T strands.}
\label{p3t3_free}
\end{center}
\end{figure}

\begin{figure}
\begin{center}
\includegraphics[width=100pt]{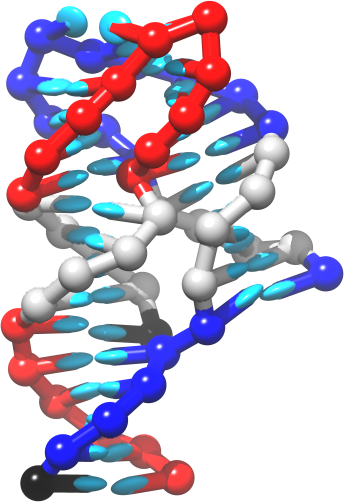}
\caption{(Color online) Example configuration of a metastable kissing complex, found from the FFS simulations of P$_4$T$_4$. The black nucleotides indicate the 3$^\prime$ end of the strand, while the gray nucleotides denote intact 4-stem hairpins. In total there are 15 correctly aligned base pairs, 5 connect the tails of the hairpins while 10 base pairs connect the loops of the hairpins. }
\label{pseudoknot}
\end{center}
\end{figure}

\begin{figure}
\begin{center}
\includegraphics[width=225pt]{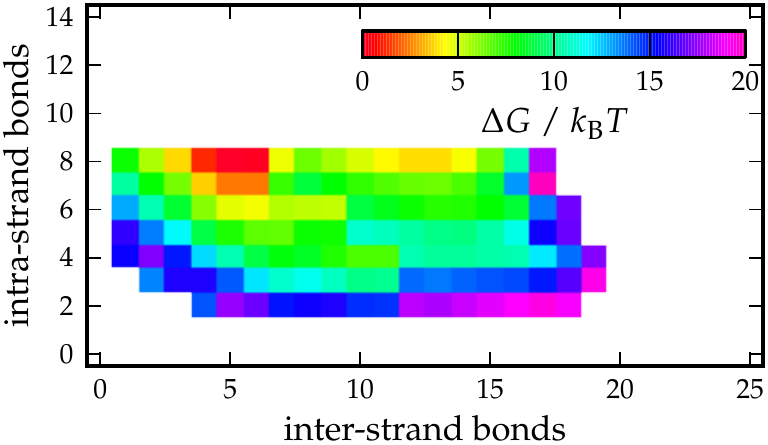}
\caption{(Color online) Free-energy profile of the metastable states in the P$_4$T$_4$ system. The free energy in both figures is plotted as a function inter- and intra-strand base pairs, where each free energy was measured relative to the state (5,8).} 
\label{p4t4_free}
\end{center}
\end{figure}

Free-energy landscapes for the P$_0$T$_0$ and P$_3$T$_3$ systems are plotted in Fig~\ref{p3t3_free}, while the profile for P$_4$T$_4$ was shown in Fig.~\ref{free_energy}(b) in the main text. According to Figs.~\ref{p3t3_free}(a), strands that have formed one inter-strand base pair may contain a variety of intra-strand base pairs, largely reflecting the fact that the unbound P$_0$ and T$_0$ strands were found to contain some transient intra-strand base pairs. 
There are no obvious partially-bound metastable states in this landscape.

In the P$_3$T$_3$ case, the hairpins are relatively stable and their effects on the free-energy landscape are more apparent than were the effects of secondary structure on the P$_0$T$_0$ system. In particular, significant hairpin structure is likely to be present in one or both strands when one inter-strand bond exists between the two strands, while the strands  lacking secondary structure but containing one inter-strand base pair are significantly less probable. As was pointed out in the main article, the barrier separating bound from unbound states is about 0.6 $k_B T$ higher than it was for P$_0$T$_0$, which is caused by the hairpins sequestering bases that reduces the number of available binding sites. 

Up until $\sim$13 inter-strand base pairs, the states with the lowest free-energy all contain hairpin structure. As described in the main article, if the system is to increase the number of base pairs, one or both hairpins must melt. When the hairpins do begin to melt, the free energy for these states largely converges to the free energy of the comparable P$_0$T$_0$ structure containing the same number of base pairs (this is also illustrated in Fig. 2(a) of the main article). In this landscape, there is some evidence of barriers between intermediate states with and without hairpins, consistent with the existence of metastable states.
 
Finally, the P$_4$T$_4$ free-energy profile was discussed in the main article. We also computed separately the free energies of the intermediate states relative to the state (5,8), which is shown in Fig.~\ref{p4t4_free}, using an order parameter designed to allow the sampling of the pseudoknotted intermediate with inter-strand base pairs between stem and loop. The plot in the main text clearly shows a metastable intermediate at  (5,8), corresponding to two hairpins bound by their tails. This is also visible in Fig.~\ref{p4t4_free}, as is a second local minimum at (13,8) corresponding to the pseudknotted state. This plot indicates that the pseudoknotted state is less stable than the tail-only state. Kinetic results in Section~\ref{kinetics_details} show that interchange between the two minima is reasonably fast.

\subsection{Kinetics}
\label{kinetics_details}

The hybridization rate constants, $k_+$, the melting rate constants, $k_-$, and the equilibrium constants $K_{eq}$, for each system are listed in Table~1 of the main article. The melting rates, $k_-$, were not computed using FFS, but rather by using Eq.~(1) in the main article combined with our calculations of $k_+$ and $\exp(\Delta G^0 / k_B T)$ from the free-energy calculations in Section~\ref{duplex_free_energies}. The cumulative statistics of the FFS simulations for P$_0$T$_0$, P$_3$T$_3$, and P$_4$T$_4$ systems are presented in Tables~\ref{p0t0_ffs},~\ref{p3t3_ffs} and~\ref{p4t4_ffs}, respectively. 


\begin{table*}
    \renewcommand{\arraystretch}{1.5}
    \begin{tabular}{>{\centering}m{1cm} >{\centering}m{2cm} >{\centering}m{3cm} c }
    \hline
    \multicolumn{4}{c}{P$_0$T$_0$} \\
    \hline
    $\lambda$ & Crossings & Total time & Flux \\
    \hline
    $\lambda_{-1}^{0}$ & 50212 & \num{1.09e-03}\SI{}{\second} & \num{4.33e+07} $\pm$ \num{2.26e+05}\SI{}{\per\second} \\

    $\lambda$ & Success & Attempts & Fractional success \\
    \hline
    $\lambda_{0}^{1}$ & 25000 & 78436 & 0.319 $\pm$ 0.002 \\
    $\lambda_{1}^{2}$ & 24900 & 116474 & 0.214 $\pm$ 0.004  \\
    $\lambda_{2}^{3}$ & 25000 & 489155 & 0.051 $\pm$ 0.002 \\
    $\lambda_{3}^{4}$ & 5000 & 36617 & 0.139 $\pm$ 0.010  \\
    $\lambda_{4}^{5}$ & 5000 & 7856 & 0.644 $\pm$ 0.036 \\
    $\lambda_{5}^{6}$ & 5000 & 5001 & 0.999 $\pm$ 0.0001  \\
    $\lambda_{6}^{7}$ & 500 & 500 & 1.000 $\pm$ 0.000  \\
    \hline
    \end{tabular}
    \caption{Results of FFS for the hybridization of P$_0$ and T$_0$ strands. The flux was measured for the crossing of $\lambda_{-1}^{0}$ and probabilities of reaching $\lambda_{Q-1}^{Q}$ from $\lambda_{Q-2}^{Q-1}$. }
    \label{p0t0_ffs}
\end{table*}

\begin{table*}
    \renewcommand{\arraystretch}{1.5}
    \begin{tabular}{>{\centering}m{1cm} >{\centering}m{2cm} >{\centering}m{3cm} c }
    \hline
    \multicolumn{4}{c}{P$_3$T$_3$} \\
    \hline
    $\lambda$ & Crossings & Total time & Flux \\
    \hline
    $\lambda_{-1}^{0}$ & 50227 & \num{1.35e-03}\SI{}{\second} & \num{3.68e+07} $\pm$ \num{2.07e+05}\SI{}{\per\second} \\

    $\lambda$ & Success & Attempts & Fractional success \\
    \hline
    $\lambda_{0}^{1}$ & 25000 & 88757 & 0.282 $\pm$ 0.001  \\
    $\lambda_{1}^{2}$ & 25000 & 125327 & 0.200 $\pm$ 0.003  \\
    $\lambda_{2}^{3}$ & 25000 & 638923 & 0.039 $\pm$ 0.002 \\
    $\lambda_{3}^{4}$ & 5000 & 26560 & 0.199 $\pm$ 0.020  \\
    $\lambda_{4}^{5}$ & 5000 & 12454 & 0.410 $\pm$ 0.031 \\
    $\lambda_{5}^{6}$ & 4980 & 5152 & 0.967 $\pm$ 0.006  \\
    $\lambda_{6}^{7}$ & 500 & 500 & 1.000 $\pm$ 0.000  \\
    \hline
    \end{tabular}
    \caption{Results of FFS for the hybridization of P$_3$ and T$_3$ strands. The flux was measured for the crossing of $\lambda_{-1}^{0}$ and probabilities of reaching $\lambda_{Q-1}^{Q}$ from $\lambda_{Q-2}^{Q-1}$.} 
    \label{p3t3_ffs}
\end{table*}

\begin{table*}
    \renewcommand{\arraystretch}{1.5}
    \begin{tabular}{>{\centering}m{1cm} >{\centering}m{2cm} >{\centering}m{3cm} c }
    \hline
    \multicolumn{4}{c}{P$_4$T$_4$} \\
    \hline
    $\lambda$ & Crossings & Total time & Flux \\
    \hline
    $\lambda_{-1}^{0}$ & 50227 & \num{1.46e-03}\SI{}{\second} & \num{3.49e+07} $\pm$ \num{7.27e+05}\SI{}{\per\second} \\

    $\lambda$ & Success & Attempts & Fractional success \\
    \hline
    $\lambda_{0}^{1}$ & 25000 & 77246 & 0.324 $\pm$ 0.003  \\
    $\lambda_{1}^{2}$ & 25000 & 113764 & 0.220 $\pm$ 0.002  \\
    $\lambda_{2}^{3}$ & 25000 & 856983 & 0.030 $\pm$ 0.002 \\
    $\lambda_{3}^{4}$ & 5000 & 23011 & 0.242 $\pm$ 0.037  \\
    $\lambda_{4}^{5}$ & 5000 & 16251 & 0.336 $\pm$ 0.050 \\
    $\lambda_{5}^{6}$ & 4720 & 21786 & 0.232 $\pm$ 0.025  \\
    $\lambda_{6}^{7}$ & 5066 & 5869 & 0.864 $\pm$ 0.014  \\
    \hline
    \end{tabular}
    \caption{Results of FFS for the hybridization of P$_4$ and T$_4$ strands. The flux was measured for the crossing of $\lambda_{-1}^{0}$ and probabilities of reaching $\lambda_{Q-1}^{Q}$ from $\lambda_{Q-2}^{Q-1}$. }
    \label{p4t4_ffs}
\end{table*}

%

\subsubsection{Considerations of the Kinetic Intermediate States}
\label{subtleties}

The formation of the P$_4$T$_4$ duplex is suggested by Gao {\it et al.} to have two different kinetic regimes which cannot be fitted to a simple two-state model. They propose  a `fast' regime where the tails of the hairpins bond (or perhaps their loops kiss); and a `slow' regime where the hairpin stems are displaced by inter-strand base pairs as the strands zip up. The fast regime has a rate constant smaller by a factor of 6, and the slow regime has a rate constant smaller by a factor of 25, than the P$_0$T$_0$ duplex. In order to see non-second-order behavior, the intermediate state needs to be long-lived so that {\it both} dissociation and completion of the reaction are slow relative to the association rate. 

\begin{figure*}
\begin{center}
\includegraphics[width=470pt]{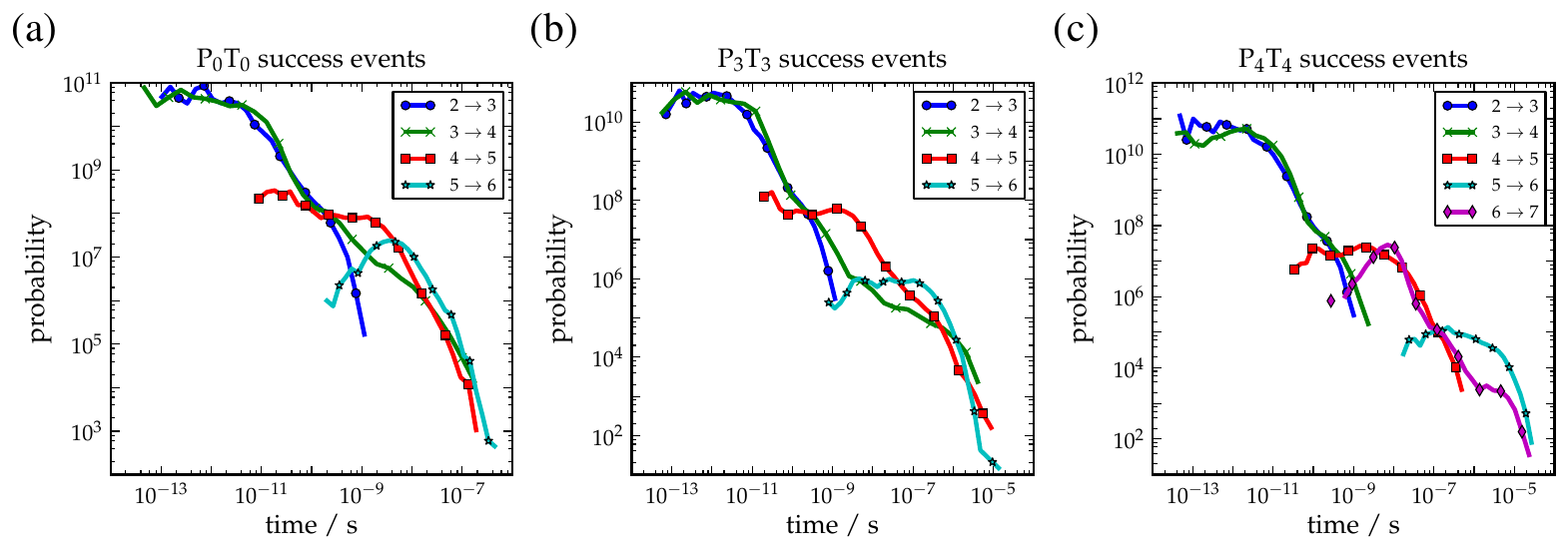}
\end{center}
\caption{Histograms of the computed rearrangement times for configurations have started from $\lambda_{Q-2}^{Q-1}$ and crossing $\lambda_{Q-1}^{Q}$ are plotted for (a) P$_0$T$_0$, (b) P$_3$T$_3$, and (c) P$_4$T$_4$. In the legend in each figure, the labels indicate the crossing of a particular interface $\lambda_{Q-1}^Q$. For example, $2 \rightarrow 3$ refers to the configuration having crossed interface $\lambda_{2}^3$ that started from $\lambda_{1}^2$. The quantity plotted on the $y$-axis is actually a probability density. Note that the times for configurations that crossed $\lambda_{6}^7$ coming from $\lambda_{5}^6$ are not plotted as they were found to be neglible.}
\label{success_times}
\end{figure*}

\begin{figure*}
\begin{center}
\includegraphics[width=470pt]{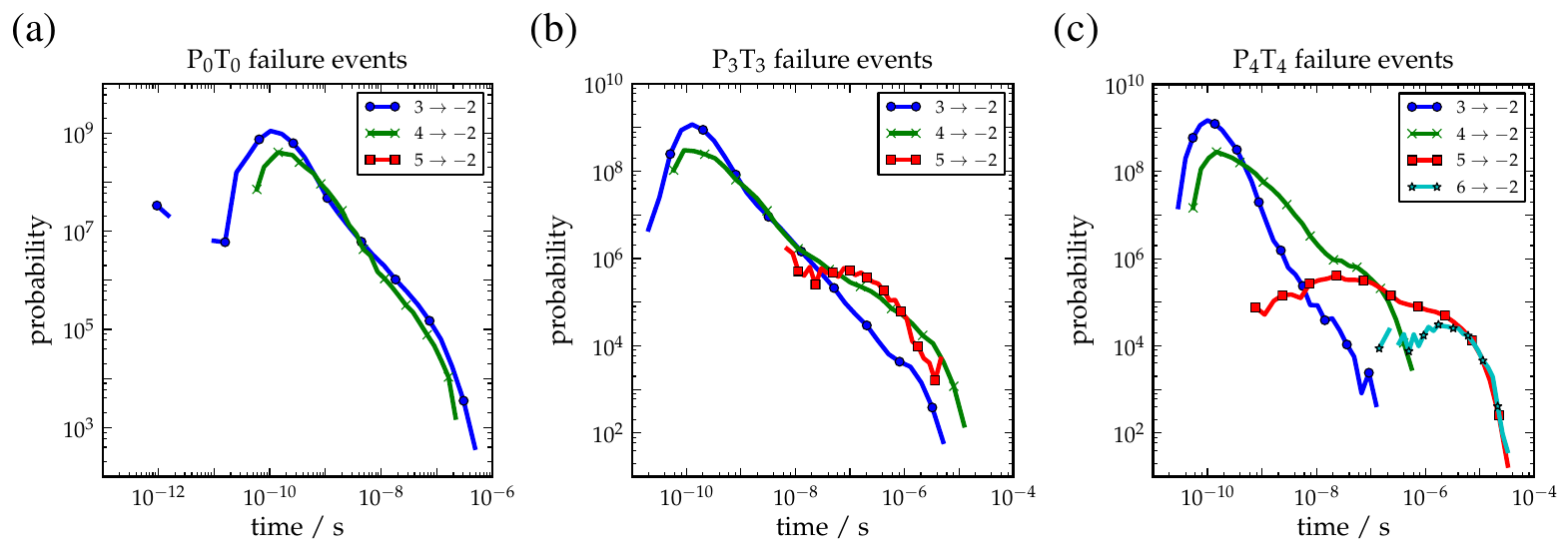}
\end{center}
\caption{Histograms of the computed rearrangement times for configurations starting from $\lambda_{Q-1}^{Q}$ and crossing $\lambda_{-2}^{-1}$ before crossing $\lambda_{Q}^{Q+1}$ are plotted for (a) P$_0$T$_0$, (b) P$_3$T$_3$, and (c) P$_4$T$_4$. Similar to Fig.~\ref{success_times}, the labels in each legend indicate a configuration having crossed a particular interface $\lambda_{Q-1}^{Q}$. For example, $6 \rightarrow -2$ refers to a configuration having intially crossing $\lambda_{5}^{6}$, but crossed $\lambda_{-2}^{-1}$ before crossing $\lambda_{6}^{7}$. As in Fig.~\ref{success_times}, the quantity plotted on the $y$-axis is actually a probability density. Note that the times for configurations that crossed $\lambda_{-2}^{-1}$ coming from $\lambda_{5}^6$ are not plotted as they were found to be neglible.}
\label{falloff_times}
\end{figure*}

We find the rearranging time for metastable states to proceed to a full duplex for the majority of simulations to be typically less than $\sim$\num{1e-7} seconds for P$_0$T$_0$ (Fig.~\ref{success_times}(a)), and $\sim$\num{1e-5} seconds for P$_3$T$_3$ and P$_4$T$_4$ (Figs.~\ref{success_times}(b) and (c)), while the longest rearrangement times that lead to dissociation events took less than $\sim$\num{1e-5} for P$_3$T$_3$ (Fig.~\ref{falloff_times}(b)) and $\sim$\num{4e-4} seconds for P$_4$T$_4$ (Fig.~\ref{falloff_times}(b)). Comparing the longest rearrangement and disassociation times with the diffusion time, we find for all duplex systems studied that the most long-lived metastable states (in P$_3$T$_3$ and P$_4$T$_4$ systems) have a lifetime slightly smaller than the rate at which they are produced. Thus reactions are second order to a reasonable approximation in oxDNA at concentrations of \SI{42}{\micro\Molar} (justifying our simulation procedure). Further, our model, consistent with Nupack, both suggest that regardless of completion rate, the 6-base pair toeholds ({\it i.e.} the tails of the intended 4-stem hairpins) are not stable enough to give rise to two kinetic regimes at the experimental concentration used. The strands just fall off too quickly, which is unsurprising given the known physics of DNA. We find that the kissing hairpin loops are even less stable, as is the pseudoknotted configuration formed when both the loops and tails bind.

Our simulations do not support the claim of two different kinetic regimes for the formation of the P$_4$T$_4$ duplex. Therefore, non-second-order behavior in Gao's experiment, if real, must be due to some unknown aspect of DNA thermodynamics that is not incorporated into the oxDNA model. 

\subsubsection{Success Probabilities of Initial Base Pairs }

\begin{figure*}
\begin{center}
\includegraphics[width=2\columnwidth]{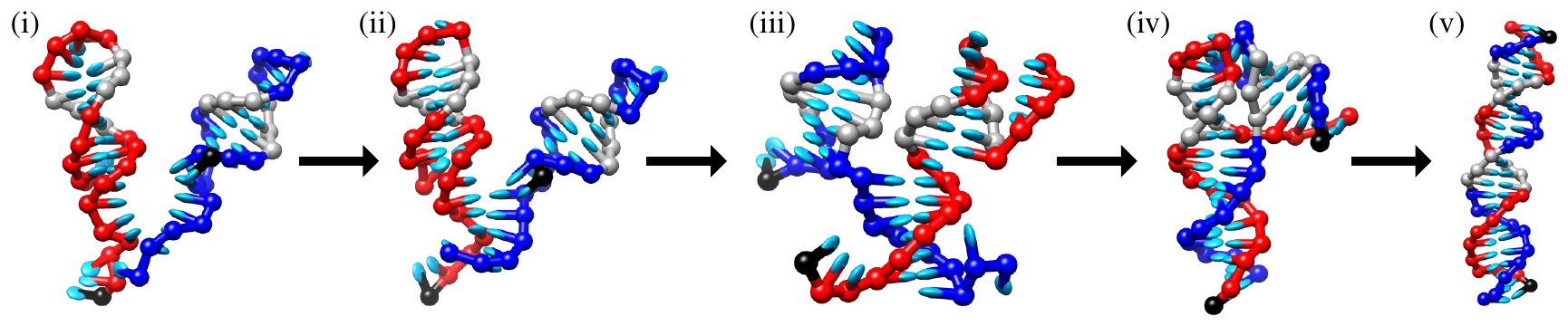}
\end{center}
\caption{(Color online) In this P$_3$T$_3$ pathway, both intended 3-base pair hairpin stems are intact just before and after the association event, seen in (i) and (ii), respectively, and are still present when the system contains 6 (iii) and 15 (iv) inter-strand base pairs. The latter configuration contains 10 base pairs between the hairpin stems and 5 base pairs between the hairpin loops. Panel (v) shows the complete duplex after the hairpins have melted. The black nucleotide indicates the 3$^\prime$ end of each strand. Gray nucleotides indicate the locations of the intended 3-stem hairpins present during the association event.}
\label{p3t3_pathway}
\end{figure*}

\begin{figure*}
\begin{center}
\includegraphics[width=2\columnwidth]{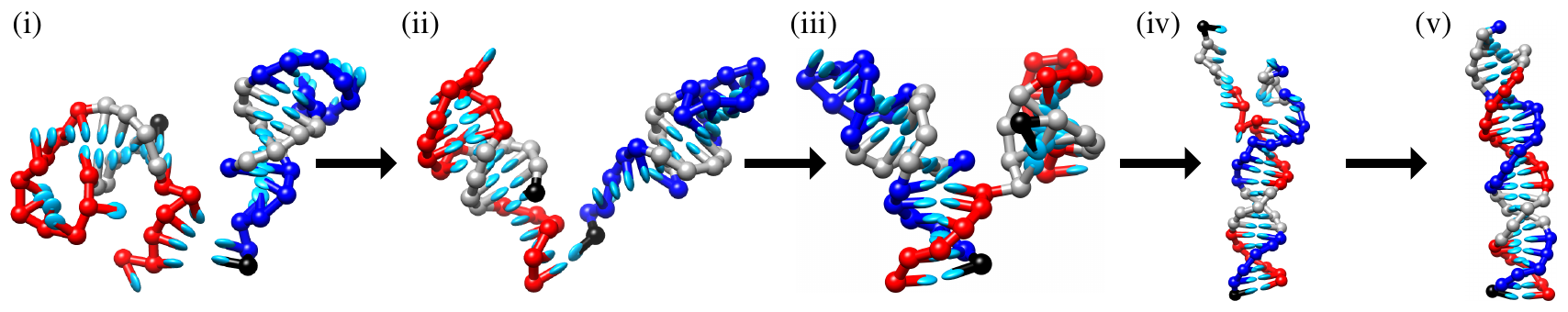}
\end{center}
\caption{(Color online) A typical duplex formation pathway for P$_4$ and T$_4$. Strands both initially possessing hairpins with 4-base pair stems are shown in (i) and (ii) just before and after the association event. (ii) The tails of the hairpins form up to 6 base pairs while retaining all base pairs in the stems. (iii) The hairpin stems have melted allowing the strands to zipper up to 15 base pairs. (v). Fully formed duplex. The black nucleotide indicates the 3$^\prime$ end of each strand. Gray nucleotides indicate the locations of the intended 4-stem hairpins present during the association event.
}
\label{p4t4_pathway}
\end{figure*}

In the top panel of Figs.~\ref{freq_2}-\ref{freq_5} we plot the frequency that a given base in the P strand is involved in base pairing for configurations that have crossed interfaces $\lambda_{Q-1}^Q$, for $Q=3,\dots,6$, respectively. In the bottom panel of Figs.~\ref{freq_2}-\ref{freq_5} we plot the probability that the base pairs between the strands led to a fully-bound duplex state, given that a configuration crossed the $\lambda_{Q-1}^Q$ interface, for $Q=3,\dots,6$, respectively. The results for the interface $\lambda_2^3$ were discussed in the main article. Here, we point out other details concerning the effects the hairpins have on the hybridization rates that have not yet been mentioned. Fig.~\ref{contact_and_success}(a) in the main text showed that the P$_0$ T$_0$ strands could be initially bound at nearly any site, but GC base pairs were more likely to lead to a duplex. As more base pairs form, the probability of eventual success increases, and is effectively unity at $\lambda_5^6$ when at least 6 base pairs has formed. 

Unlike in the P$_4$T$_4$ case where two 4-base pair hairpins dominated, there may be different combinations of 3-base pair hairpins present in the P$_3$ and T$_3$ strands during association, as was discussed in Section~\ref{strand_free}.  Because of this, the plots for P$_3$ and T$_3$ shown in Figs.~\ref{freq_2}(b) are less revealing of the effects the hairpins had on hybridization rates than the equivalent plots for P$_4$T$_4$. Overall, Fig.~\ref{freq_2}(b) illustrates that initial base pairs between P$_3$ and T$_3$ strands could occur at any binding site, and successful binding events could come from any initial base pair, but there seems to be a bias away from the center of the strand because sites 10-12 (gray color in Fig.~\ref{freq_2}(b)) may form base pairs with the bases at sites 1-3 (yellow color in Fig.~\ref{freq_2}(b)) or 18-20 (gray color in Fig.~\ref{freq_2}(b)). 

Fig.~\ref{P3T3_pathway} of the main article illustrated a successful trajectory where the P strand contained an intended 3-base pair hairpin, while the T strand contained an alternative 3-base pair hairpin. The figure also showed that hairpins were present when configurations had crossed $\lambda_4^5$, but had melted once the $\lambda_5^6$ interface had been crossed. By contrast in Fig.~\ref{p3t3_pathway}, both strands contain the intended 3-base pair hairpins during association, and were still present once the system had crossed the $\lambda_5^6$ interface. The configuration in Fig.~\ref{p3t3_pathway}(iv) has 10 base pairs between the tails and 5 base pairs between the loops of the hairpins (analogous to the pseudoknotted structure observed for  P$_4$T$_4$ depicted in Fig.~\ref{pseudoknot}). However, in this case it is extremely unlikely the two strands would disassociate before the hairpins melt, which is a relatively fast process for 3-base pair hairpins.

Turning our attention to other features of the P$_4$T$_4$ reaction, Fig.~\ref{contact_and_success} in the main article showed that initial base pairs between the two strands were found to occur between the tails or loops of the hairpins, but that successes overwhelmingly came from tail-tail initial base pairs. The probability of successful bindings leading to a duplex was also diminished by about a factor of 5 compared to P$_0$T$_0$ and P$_3$T$_3$ initial binding success probabilities. 
Once the system crosses the $\lambda_4^5$ interface, there is still 80\% probability that the system will dissociate, as is illustrated in Fig.~\ref{freq_4}(c), and is due to increased stability of the hairpins and the reduced stability of the metastable states prior to hairpin disruption in P$_4$T$_4$. Table~\ref{p4t4_ffs} and Fig.~\ref{freq_5}(c) show that even systems that cross the $\lambda_5^6$ interface still have a 15\% probability of dissociation, and base pairs are rarely present between the 3$^\prime$ end of the P strand and the 5$^\prime$ end of the T strand. We observed that pseudoknotted intermediate states such as that in Fig.~\ref{pseudoknot} were responsible for the failure events that occurred after the system crossed the $\lambda_5^6$ interface. When structures dissociate from these states, either the base pairs between the tails or between the loops melt first, which is then followed by further melting of the remaining base pairs. 

In Fig.~\ref{p4t4_pathway} a typical successful trajectory is illustrated for the P$_4$T$_4$ system, and shows that the hairpins are present when the tails are initially bound (Fig.~\ref{p4t4_pathway}(ii)), until 6 base pairs between the tails have formed (Fig.~\ref{p4t4_pathway}(iii)) when they interfere with the zippering up of the strands. Once one hairpin has opened, the other may be opened by thermal melting or strand-displacement (Fig.~\ref{p4t4_pathway}(iv) shows the scenario where the hairpins have melted but the duplex is not yet fully formed). Once freed of hairpin structure, the strands proceed to zipper up into a full duplex (Fig.~\ref{p4t4_pathway}(v)).

\subsubsection{Effects Due to Mis-aligned Inter-strand Base Pairs}

\begin{figure*}
\begin{center}
\includegraphics[width=470pt]{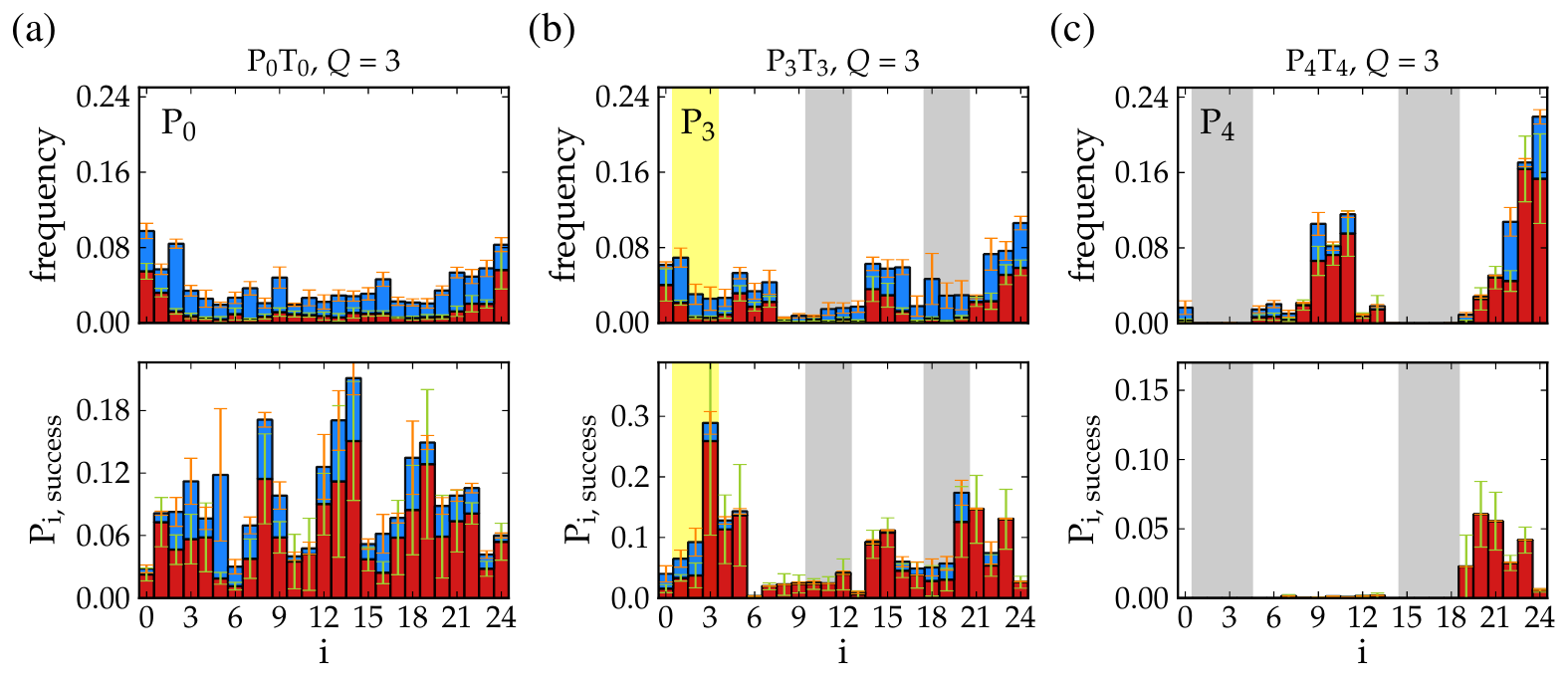}
\end{center}
\caption{(Color online) The frequencies of attachment locations for configurations that crossed $\lambda_2^3$ (labeled state $Q = 3$ in the plots) as a function of base pair index on the P strand are plotted in the top panel in (a) for P$_0$T$_0$, in (b) for P$_3$T$_3$, and in (c) for P$_4$T$_4$. The probability that said base pairs lead to a duplex are plotted in the bottom panels in (a) for P$_0$T$_0$, in (b) for P$_3$T$_3$, and in (c) for P$_4$T$_4$. In each panel, red color indicates contributions from correctly aligned duplex base pairs between the two strands, while blue color indicates contributions from misaligned base pairs between the two strands. The yellow region indicates the location of a non-intended 3-base pair hairpin, which pairs with bases at locations 10-12. The grayed out regions for P$_3$ indicate the intended 3-base pair hairpin stem, while the grayed out regions for P$_4$ indicate the 4-base pair hairpin stem. For P$_4$T$_4$ sites 5-14 are within the loop of the hairpin, and sites 19-24 are a dangling tail.}
\label{freq_2}
\end{figure*}

\begin{figure*}
\begin{center}
\includegraphics[width=470pt]{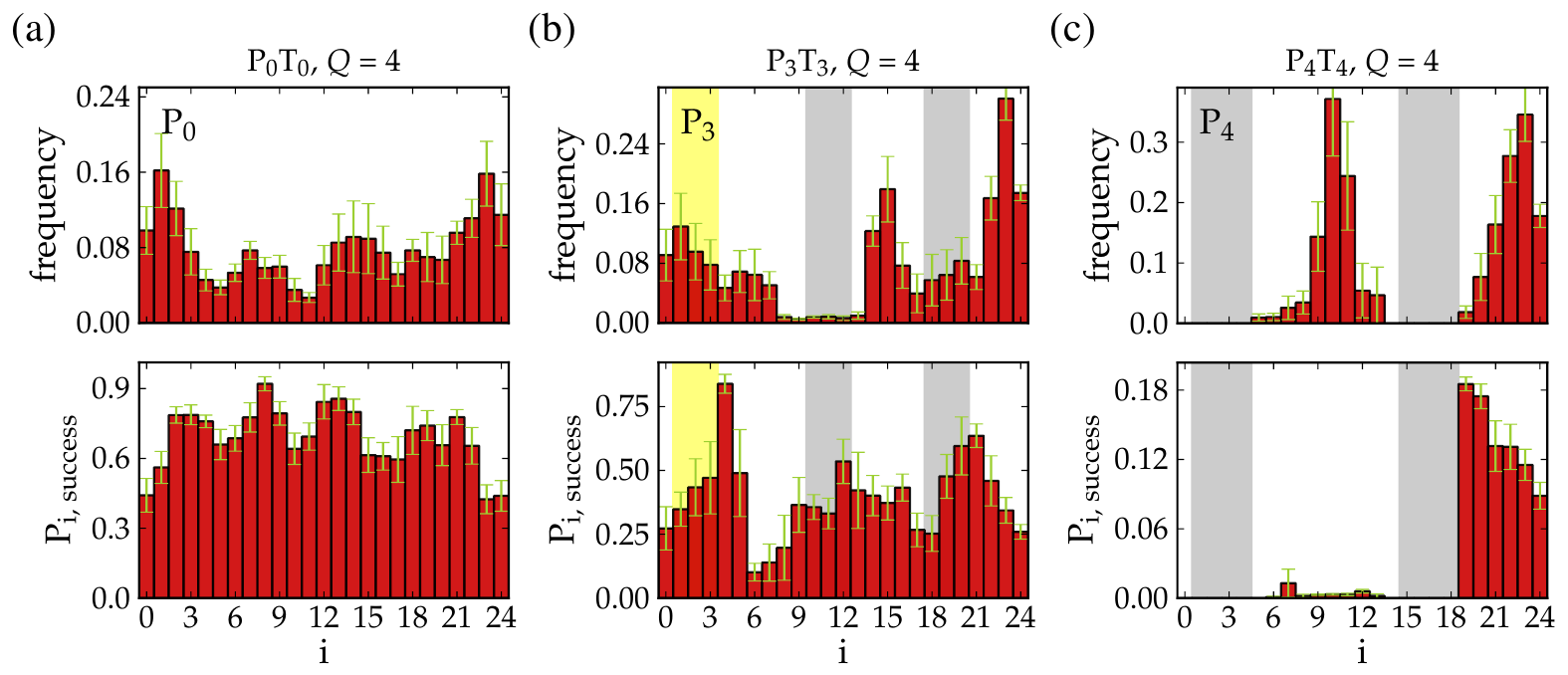}
\end{center}
\caption{(Color online) The frequencies of attachment locations for configurations that crossed $\lambda_3^4$ (labeled state $Q = 4$ in the plots) as a function of base pair index on the P strand are plotted in the top panel in (a) for P$_0$T$_0$, in (b) for P$_3$T$_3$, and in (c) for P$_4$T$_4$. The probability that said base pairs lead to a duplex are plotted in the bottom panels in (a) for P$_0$T$_0$, in (b) for P$_3$T$_3$, and in (c) for P$_4$T$_4$. Similar to Fig.~\ref{freq_2}, the yellow region indicates a the location of a non-intended 3-base pair hairpin, which pairs with bases at locations 10-12. The grayed out regions for P$_3$ indicate the intended 3-base pair hairpin stem, while the grayed out regions for P$_4$ indicate the 4-base pair hairpin stem. Sites 5-14 are within the loop of the hairpin, and sites 19-24 are a dangling tail.}
\label{freq_3}
\end{figure*}

\begin{figure*}
\begin{center}
\includegraphics[width=470pt]{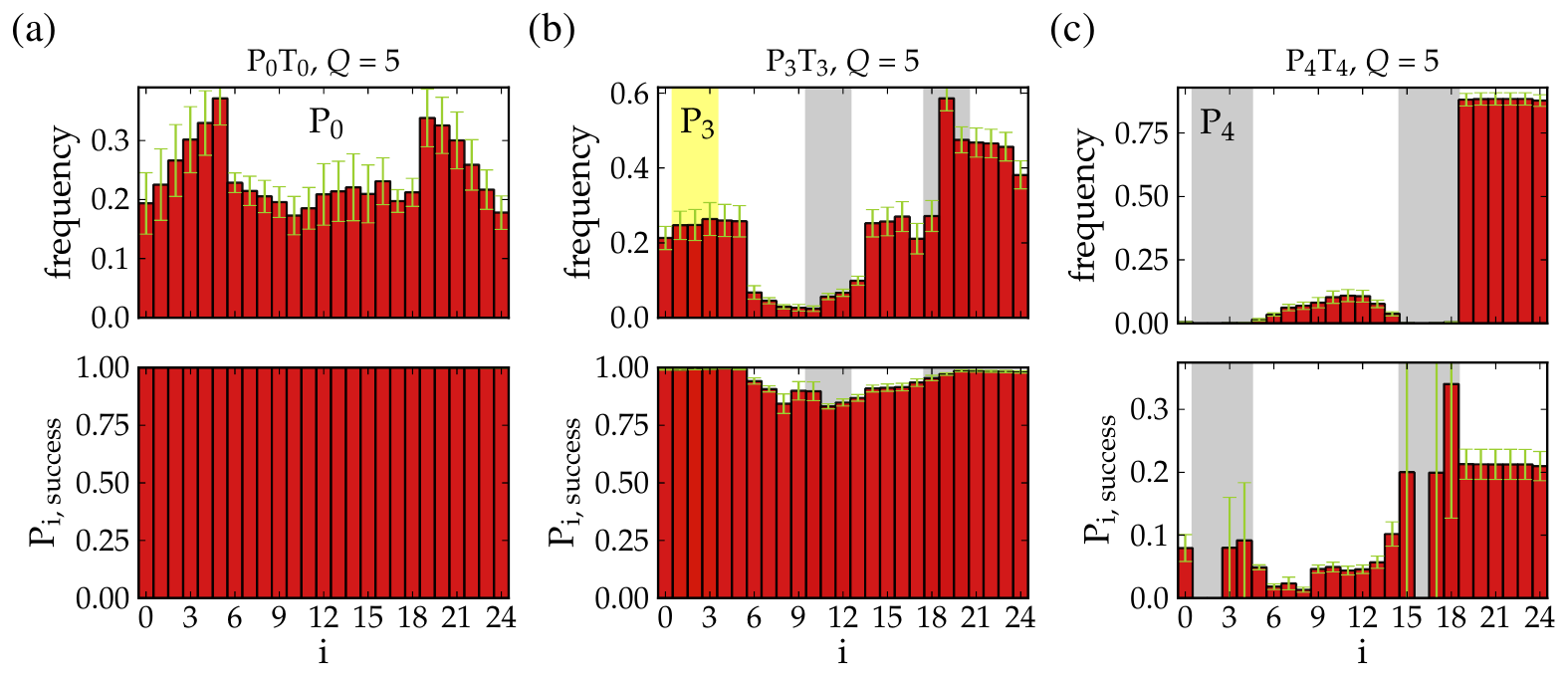}
\end{center}
\caption{(Color online) The frequencies of attachment locations for configurations that crossed $\lambda_4^5$ (labeled state $Q = 5$ in the plots) as a function of base pair index on the P strand are plotted in the top panel in (a) for P$_0$T$_0$, in (b) for P$_3$T$_3$, and in (c) for P$_4$T$_4$. The probability that said base pairs lead to a duplex are plotted in the bottom panels in (a) for P$_0$T$_0$, in (b) for P$_3$T$_3$, and in (c) for P$_4$T$_4$. Similar to Fig.~\ref{freq_2}, the yellow region indicates a the location of a non-intended 3-base pair hairpin, which pairs with bases at locations 10-12. The grayed out regions for P$_3$ indicate the intended 3-base pair hairpin stem, while the grayed out regions for P$_4$ indicate the 4-base pair hairpin stem. Sites 5-14 are within the loop of the hairpin, and sites 19-24 are a dangling tail.}
\label{freq_4}
\end{figure*}

\begin{figure}
\begin{center}
\includegraphics[width=150pt]{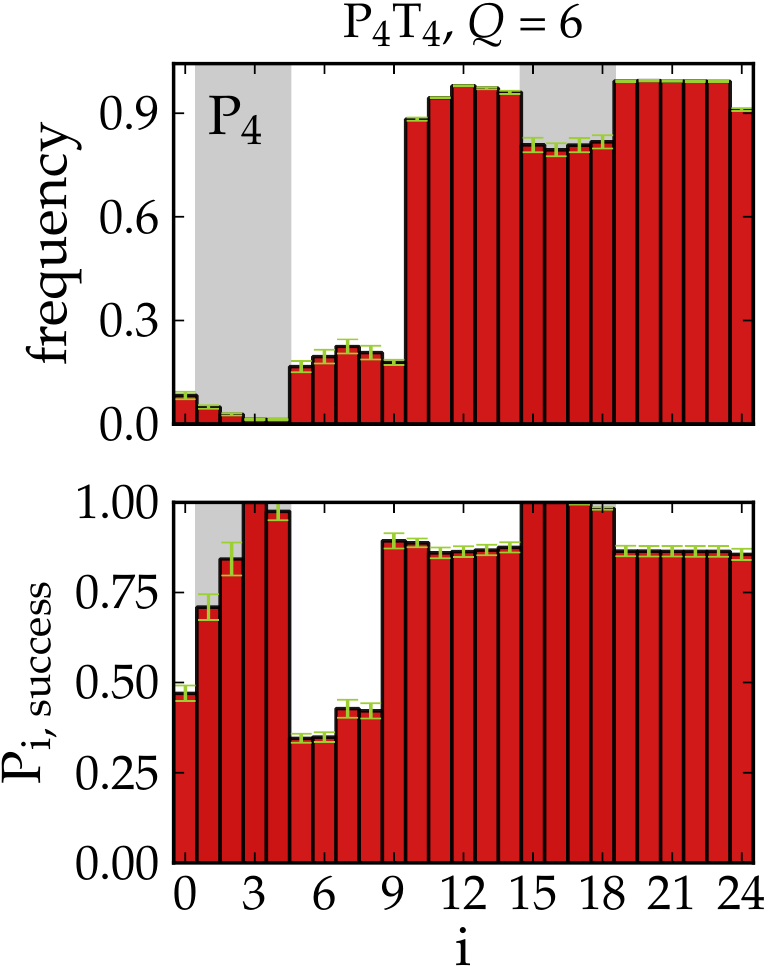}
\end{center}
\caption{(Color online) For P$_4$T$_4$, the frequencies of attachment locations for configurations that crossed $\lambda_5^6$ (labeled state $Q = 6$ in the plots), as a function of base pair index on the P strand, are plotted in the top panel, and the probability that said base pairs lead to a duplex are plotted in the bottom panel. The grayed out regions for P$_4$ indicate the 4-base pair hairpin stem. Sites 5-14 are within the loop of the hairpin, and sites 19-24 are a dangling tail.}
\label{freq_5}
\end{figure}

Fig.~\ref{freq_2}(a) and (c) are slightly different from Fig.~\ref{contact_and_success} in the main article. In this case the initial association base pairs, and their probabilities of success, have been separated into contributions from correctly aligned base pairs, and mis-aligned base pairs. Tables~\ref{p3t3_ffs}-\ref{p4t4_ffs} tabulate the success probabilities between interfaces as hybridization proceeds. The probabilities of crossing the $\lambda_3^4$ interface in the simulations clearly shows that the P$_0$T$_0$ system is less successful than the P$_3$T$_3$ and P$_4$T$_4$ systems at forming subsequent base pairs after the first has formed, where P$_0$T$_0$, P$_3$T$_3$ and P$_4$T$_4$ systems have a 14\%, 20\%, and 24\% chance of successfully crossing the $\lambda_{3}^{4}$ interface. The effect is due to the higher incidence of mis-aligned base pairs forming compared to correctly aligned base pairs (see Fig.~\ref{freq_2}, top panel). Mis-aligned base pairs are more likely to break, resulting in strand separation, as opposed to leading to more inter-strand base pairs. Furthermore, the P$_3$T$_3$ system is more likely to form mis-aligned base pairs than the P$_4$T$_4$ system. Generically, two strands containing $N$ binding sites, which contain negligible secondary structure, can be bound in $\sim N^2$ number of possible ways. For P$_3$T$_3$ and P$_4$T$_4$ systems there are less ways for the strands to mis-align during association due to larger hairpins present in the two systems obscuring binding sites. This is evident from the bottom panels in Fig.~\ref{freq_2}, which shows a decreasing likelihood that mis-aligned base pairs lead to a full duplex. For the P$_4$T$_4$ system, mis-aligned base pairs almost never lead to a full duplex, but rather they lead to strand separation. The increased probability of P$_3$T$_3$ configurations crossing $\lambda_3^4$ compared to P$_0$T$_0$ is offset by a lower probability of crossing $\lambda_4^5$ (because of the blocking effects of the hairpins). The combination of these two effects gives rise to very similar probabilities of achieving a full duplex given 1 base pair for P$_0$T$_0$ and P$_3$T$_3$, as noted in Table~1 in the main text.

\end{document}